%% file: main.tex
\documentclass[ASNA,twocolumn]{USG} %twocolumn

\usepackage[numbers]{NJDnatbib}
\usepackage{amssymb,algorithm,algorithmicx}
\usepackage{anyfontsize} %

\usepackage{colortbl}   % For coloring table lines
\usepackage{xcolor}     % For color definitions

\usepackage{steinmetz}

\definecolor{AgClaude}{HTML}{7375B3}   % slate
\definecolor{AgCodex}{HTML}{4AAE8D}   % teal
\definecolor{AgGemini}{HTML}{DA764D}  % orange
\definecolor{LPy}{HTML}{4A7AB5}
\definecolor{LJava}{HTML}{D94040}
\definecolor{LJS}{HTML}{2BAF6A}
\definecolor{LCpp}{HTML}{C8A020}
\definecolor{HeadTint}{HTML}{ECECF5}  % light slate for header rows
\definecolor{SumTint}{HTML}{E4F2EC}   % light teal for summary blocks
\usepackage{enumitem}
\usepackage{tikz}
\usetikzlibrary{shapes.geometric, arrows.meta, positioning, calc}

\SetEnumitemKey{itemnosep}{noitemsep}

\articletype{Article}%

\journal{AI Magazine}
\volume{0}
\copyyear{2026}
\startpage{1}
\articledoi{0000}

\begin{document}
\title{Code That Works, Environments That Don't: Measuring Environment Reproducibility in AI-Generated Software}
%\subtitle{Subtitle}

\author[1]{Bhanu Prakash Vangala}
\author[1]{Tanu Malik}
\authormark{Vangala \textsc{et al.}}
\titlemark{Environment Specification Gap in LLMs}
\address[1]{\orgdiv{Department of Electrical Engineering and Computer Science}, \orgname{University of Missouri--Columbia}, \orgaddress{\state{Missouri}, \country{USA}}}
\corres{Bhanu Prakash Vangala (\email{bv3hz@missouri.edu}); Tanu Malik (\email{tanu@missouri.edu})}

\setcounter{secnumdepth}{3}

\input{sections/00_abstract}

\maketitle

\input{sections/01_introduction}
\input{sections/02_protocol}
\input{sections/03_primitives}
\input{sections/04_setup}
\input{sections/05_metrics}
\input{sections/06_results}
\input{sections/07_related}
\input{sections/08_conclusion}

%\input{sections/methodology_old}

\bibliography{references}

\input{sections/09_appendix}

\end{document}

%% file: sections/00_abstract.tex
% ═══════════════════════════════════════════════════════════════════════
% ABSTRACT
% In the Wiley USG class the abstract is a preamble command:
%   \abstract[ABSTRACT]{ ... }
% placed BEFORE \maketitle. The original \begin{abstract}...\end{abstract}
% environment has therefore been wrapped in \abstract[ABSTRACT]{...}.
% ═══════════════════════════════════════════════════════════════════════
\abstract[ABSTRACT]{
Code generation has emerged as a central capability of large language models, with coding agents now able to produce functionally correct software projects from natural language prompts. However, functional correctness alone does not capture a critical dimension of generation quality: environment specification, which is defined as the accurate identification of the dependencies required to execute generated code, is equally critical. 
We develop an agent protocol for environment specification, and introduce a three-layer framework comprising declared , runtime-installed, and necessary and sufficient dependencies to systematically assess coding agents for environment specification. Using this protocol, we evaluate the extent to which coding agents systematically misspecify software environment dependencies, and how does this misspecification vary across three agents, four languages, and fifty programming tasks. Our results show that current coding agents exhibit systematic generalization failures along this dimension, producing dependency specifications that are inconsistent, redundant, or incomplete in ways that functional tests do not detect.  Across agents, dependency set agreement is as low as 7\% for identical tasks, and newer agents show no meaningful improvement, suggesting the failure is not resolved by scale or recency. 
The largest divergence occurs between the declared ($D^{(1)}$) and runtime ($D^{(3)}$) dependency layers, implicating environment priors learned from the models' training distributions as the primary driver. Our findings establish environment specification as a distinct, measurable axis of code generation quality that current benchmarks do not capture, and motivate training objectives and evaluation protocols that jointly optimize for functional correctness and environmental portability.
}

%% file: sections/01_introduction.tex
% ═══════════════════════════════════════════════════════════════════════
\section{Introduction}
\label{sec:intro}

Code generation has emerged as one of the most consequential capabilities of large language models. Advanced coding agents~\cite{anthropic2024claude, team2024gemini} can now synthesize complete, multi-file software repositories from natural language descriptions, representing a qualitative leap beyond the line-level autocomplete of earlier tools~\cite{chen2021codex, pearce2022asleep}. However, the benchmarks used to evaluate these agents have not kept pace with their evolving capabilities~\cite{jiang2024survey, fan2023large}.

%Code generation has emerged as one of the most consequential capabilities of large language models. Coding agents such as Claude Code, Gemini Code Assist, and Codex can now synthesize complete, multi-file software artifacts from natural language descriptions, which is a qualitative leap beyond the line-level autocomplete that characterized earlier tools~\cite{}. The standard by which these agents are evaluated has, however, not kept pace with their capabilities. 

Current evaluation frameworks~\cite{chen2021codex, austin2021mbpp, li2022alphacode} focus almost exclusively on functional code correctness. They fail to measure the correctness of the underlying software environment, such as dependencies, package versions, and system configurations, which is strictly required for execution.

One might argue that code and its environment are two sides of the same coin, implying that traditional benchmarks implicitly evaluate both. We counter that software environments are fundamentally distinct from source code. While code dictates logic, environments manage external, stateful ecosystems that are prone to silent runtime failures and complex version conflicts~\cite{decan2019empirical, wang2020empirical}. Consequently, evaluating environment generation requires a qualitatively different benchmarking paradigm. Before AI scientists build such benchmarks, we believe it is important to first systematically measure how effectively current agents specify these environments.

%Current benchmarks~\cite{} that are used to evaluate coding agents focus purely on correctness of functions specified within the code; they do not measure correctness of software environment also specified within the code. This software environment refers to dependencies and environment configuration and is necessary for reliable execution of functions in the code. One may argue that code and environment are two sides of the same coin and code-related benchmarks can be used for assessing environments. However, we argue that software environments are  fundamentally different from code necessitating qualitatively different benchmarks. Given that software environments are language specific and follow different specification standards, before developing a benchmark it is important that we measure and assess how performant are coding agents in terms of specifying software environments. 

To address this gap, this paper introduces a framework to evaluate and quantify how reliably coding agents generate software environments. We propose an iterative evaluation protocol. First, the agent is prompted to generate a complete software artifact containing both source code and its environment specification, establishing a baseline performance metric. Second, we evaluate the agent's repair capabilities by forcing it to iteratively diagnose and resolve its own environment failures. The protocol ultimately determines the agent's baseline accuracy, its self-repair efficiency, and the execution bottlenecks common to modern LLMs.

Toward this goal, this paper introduces a novel evaluation framework to measure, assess, and quantify the quality of environment specifications generated by coding agents. We propose an iterative evaluation protocol divided into two primary phases. First, an agent is prompted to generate a complete software artifact comprising both source code and its environment specification, establishing a baseline metric for its initial generation capabilities. Second, if the initial environment is under-specified or broken, the agent enters a self-repair phase where it must iteratively diagnose and resolve its own environment failures. Once an agent achieves a successfully executing artifact, the overall quality of the environment specified across all iterations is independently validated using application virtualization. By tracing the necessary and sufficient dependencies loaded into memory during runtime, our framework establishes an absolute ground truth, allowing us to explicitly measure the degree of dependency over-specification.

Our framework thus provides a holistic, three-dimensional assessment: it benchmarks initial environment generation; quantifies the repair friction required to resolve under-specification; and exposes over-specification by comparing the agent's declared dependencies against the lean, optimal runtime ground truth.

We instantiate this framework in a controlled empirical study spanning three production coding agents, four language ecosystems, and 1,000 curated project instances. Each instance is executed in a clean, isolated environment with full dependency tracing from manifest through runtime. Using this rich dataset, we systematically characterize the root causes of environment specification failures across different agents, programming languages, and task domains.

Our results produce the first systematic evidence that dependency declarations are profoundly inconsistent across agents solving identical tasks. Critically, we find that LLM environment generation is non-deterministic; a single agent rarely reproduces its own dependency choices across independent runs. In fact, the unanimous consensus rate for identical tasks across all languages is exactly zero. Furthermore, the gap between declared and actual runtime dependencies is wide, directionally biased, and heavily ecosystem-dependent. These variations implicate artifacts within the models' training distributions rather than the inherent difficulty of the tasks themselves. Our earlier study~\citep{vangala2026aigenerated} offered first evidence that agent-generated code is not reproducible at the dependency level; here we formalize the underlying evaluation protocol and measurement primitives, and expand the analysis across agents, ecosystems, and stochastic trials.
 
These findings have direct, urgent implications for the reproducibility of agent-generated software and the validity of using agents for autonomous software engineering. Specifically, our study indicates that software environment configuration must be treated as a core target training objective for LLMs. Furthermore, precision, recall, and F1-scores on dependency declarations—measured against traced runtime ground truth—must be treated as first-class evaluation criteria alongside functional code correctness.

We make the following primary contributions:

\begin{itemize}[itemnosep,leftmargin=*]
\item \textbf{Three-Layer Dependency Framework:} We introduce a formal paradigm for evaluating coding agents by decoupling dependencies into three layers: what the agent \textit{declares}, what the package manager \textit{installs}, and what the program actually \textit{loads} into memory at runtime.
\item \textbf{Comprehensive Evaluation Dataset:} We construct and open-source a dataset spanning three production coding agents, four language ecosystems, and 1,000 verified project instances, complete with execution traces in isolated environments.
\item \textbf{Empirical Analysis of Environment Failures:} We deliver a systematic characterization of LLM environment generation, proving that under- and over-specification represent systemic algorithmic gaps in training rather than isolated edge cases.
\end{itemize}

The rest of this paper is organized as follows. Section~\ref{sec:protocol} formalizes the agent evaluation protocol and defines the core agent capabilities required for environment specification. Section~\ref{sec:primitives} details the primitive functions that underpin the three-layer dependency framework. Section~\ref{sec:setup} outlines our experimental design, followed by the evaluation dimensions in Section~\ref{sec:metrics} and a comprehensive evaluation and metrics analysis in Section~\ref{sec:results}. Finally, Section~\ref{sec:related} reviews related work, and Section~\ref{sec:conclusion} concludes the paper.

%% file: sections/02_protocol.tex
\section{Agent Evaluation Protocol}
\label{sec:protocol}
%for Dependency Specification}

Given a hardware and operating environment, if a software environment is incorrectly specified, a developer will aim to fix it till it results in executable code. An agent is similar. Our agent evaluation protocol divides this agent behavior into two skills, a \textit{generation skill}, in which it produces code and a dependency manifest for a given task, and a \textit{repair skill}, in which it iteratively revises both specified code and dependency manifest in response to runtime failures. 

%To systematically evaluate a coding agent  $A$ for dependency specification, we observe that $A$ exercises two distinct skills: a \textit{generation skill}, in which it produces code and a dependency manifest for a given task, and a \textit{repair skill}, in which it iteratively revises both specified code and dependency manifest in response to runtime failures. %The former yields the initial declared dependencies, while the latter produces the final successful artifact used for installed and runtime analysis. 

Formally, given a coding agent $A$ and a programming task $T$ with corresponding prompt $\pi_{T}$, the agent's generation skill produces the initial code $C$ and dependency manifest $M$, specified as:

\begin{equation}
(C,M) = \texttt{Gen}(A, \pi_{T})
\end{equation}

%where coding agent $A$ produces full source code $C$ and  declared dependency manifest $M$.  

Let the $(C, M)$ produced by the \texttt{Gen} skill be indexed as $(C^{(0)}, M^{(0)})$. Executing $(C^{(0)}, M^{(0)})$ in an isolated environment $I$ yields an initial  signal:

$$
e^{(0)} = \texttt{Execute}(C^{(0)}, M^{(0)}, I)
$$

If $e^{(0)} = \emptyset$ the agent has reliably executed the code given the dependency manifest, and it denotes successful execution. Otherwise it denotes failed execution with $e^{(0)}$ specifying the error. If $e^{(0)} \neq \emptyset$, the \texttt{Repair} skill iteratively generates $(C^{(i)}, M^{(i)})$ by feeding the previous code, manifest, and error signal to $A$:

$$
(C^{(i)}, M^{(i)}) = \texttt{Repair}(A, I, [C^{(i-1)}, M^{(i-1)}, e^{(i-1)}]), \quad i = 1, \dots, k
$$

$$
e^{(i)} = \texttt{Execute}(C^{(i)}, M^{(i)}, I)
$$

The loop halts at the first $i \leq k$ for which $e^{(i)} = \emptyset$, or exhausts the budget $k$ otherwise. The final artifact is denoted $(C^{(k)}, M^{(k)})$, where $(C^{(k)}, M^{(k)}) = (C^{(i)}, M^{(i)})$ for the smallest $i$ at which $e^{(i)} = \emptyset$, or the last iterate if no successful execution is reached within budget.

% Let the $(C,M)$ produced by the $\text{Gen}$ skill be indexed as $(C^0, M^0)$. The repair skill $\text{Rep}$ iteratively generates $(C^i,M^i)$  by executing  $(C^{i-1}, M^{i-1})$ in an isolated environment $I$, obtaining the error if execution fails, and feeding both the error and previous code and dependency manifest to $A$. 

% \begin{equation}
% (C^{i}, M^{i}, e^{i}) = \text{Repair}(A, I, [C^{i-1}, M^{i-1}, e^{i-1}]), \quad i = 1, \dots, k
% \end{equation}

% To generate runtime dependencies, the pair $(C, M)$ is executed inside an isolated environment $I$ consisting of a base operating system (OS) image, e.g., Ubuntu 22.04 with no pre-installed packages beyond the system default. If execution fails,  $A$ applies its repair skill iteratively, with the runtime error fed back at each time step:

% % \begin{equation}
% % (C^{i},D^{i}) = Update(L, [C^{i-1},D^{i-1}], e_{i-1}),   i =  1,\ldots,k
% % \end{equation}

% where $e^{i-1}$ is the runtime error observed when executing $(C^{i-1}, M^{i-1})$ in $I$, and $k$ is a fixed iteration budget specified to $A$ via its agent configuration file. The iteration proceeds while $e^{i} \neq 0$ and halts at the first $i \leq k$ for which $e^{i} = \phi$, \textit{i.e.,} successful execution.  We denote the final successful artifact as $(C^k,M^k)$, where  $(C^k,M^k) = (C^i,M^i)$ for the smallest $i$ at which execution succeeds, or the last iterate if no successful execution is reached within budget. 

From these artifacts, obtained iteratively, we derive three dependency sets: 
\begin{itemize}[noitemsep,leftmargin=*]
\item The \textbf{declared} set $D^{(1)} := \texttt{ParseManifest}_{l}(M^{(0)})$ captures the agent's initial claims about the environment the code needs given the programming task. 
\item The \textbf{installed} set $D^{(2)}:= \texttt{Resolve}_{l}(M^{(k)},I)$ captures the environment installed in $I$ at the end of the iterative cycle. In both definitions, ${l}$ denotes the language ecosystem, which is necessary to understand the language code-specific dependencies. 
\item The \textbf{traced} set $D^{(3)} := \texttt{Trace}(C^{(k)}, I)$ is defined only when $e^{(k)} = \emptyset$, \textit{i.e.,} when the program has been successfully installed and executes without error. \texttt{Trace} is agent-independent and captures the necessary and sufficient environment for the observed execution of $C^{(k)}$ in $I$.  We often use $C^{(k)}$ to refer to both the source code and its installed executable in $I$ interchangeably, as the distinction is not material to the dependency analysis.
\end{itemize}

%\begin{equation}
%   D^{1} := Parse_{l}(M^{0}); \hspace{25pt} D^{2}:=Resolve_{l}(M^{k}); \hspace{25pt}
%   D^{3} := Trace(C^{k},I)
% $$D^{(1)} := \texttt{ParseManifest}_\ell(M^{(0)});  \hspace{25pt} D^{(2)} := \texttt{Resolve}_\ell(M^{(k)}, I)$$ %\hspace{25pt}
% $$D^{(3)} := \texttt{Trace}(C^{(k)}, I), \quad \text{defined only if } e^{(k)} = \emptyset $$
% %\end{equation}

% The \textbf{declared} set $D^{1} := Parse_{l}(M^{(0)})$ captures the agent's initial assertion about the task's requirements. The \textbf{installed} set $D^{2}:=Resolve_{l}(M^{k})$ captures all packages installed by the package manager in $I$ to successfully execute the task. In both definitions, ${l}$ denotes the language ecosystem, which is necessary to understand the code-specific dependencies. The \textbf{traced} set $D^{3} := Trace(C^{k},I)$ is produced by our evaluation pipeline via \texttt{strace}, capturing the minimal necessary and sufficient set of dependencies actually loaded during execution of  the executable corresponding to $C^{k}$ in $I$. We abuse the notation and use $C^{k}$ instead of the corresponding executable for consistency sake.  

\begin{figure}[t]
\centering
\begin{tikzpicture}[
  set/.style={ellipse, draw, minimum width=3.5cm, minimum height=1.45cm,
              font=\normalsize, align=center},
  lbl/.style={font=\scriptsize, align=center, fill=white, inner sep=2pt}
]
  %% Nodes — wider horizontal gap, taller vertical drop
  \node[set, fill=blue!12]  (D1) at (0,0)      {$D^{(1)}$\\Declared};
  \node[set, fill=teal!12]  (D2) at (5.0,0)    {$D^{(2)}$\\Installed};
  \node[set, fill=red!12]   (D3) at (2.3,-3.0) {$D^{(3)}$\\Traced};
  %% D1 <-> D2  (horizontal — label above)
  \draw[dashed, <->] (D1) -- 
    node[lbl, above=10pt]
      {repair-skill gap $\Delta$:\\
       $D^{(1)}\!\leftrightarrow\!D^{(2)}$}
    (D2);
  %% D2 <-> D3  (right leg — label to the right, shifted outward)
  \draw[dashed, <->] (D2) -- 
    node[lbl, right=6pt, pos=0.5]
      {bloat: $D^{(2)}\!\setminus\!D^{(3)}$\\
       hidden: $D^{(3)}\!\setminus\!D^{(2)}$}
    (D3);
  %% D1 <-> D3  (left leg — label to the left, shifted outward)
  \draw[dashed, <->] (D1) -- 
    node[lbl, left=6pt, pos=0.5]
      {phantom: $D^{(1)}\!\setminus\!D^{(3)}$\\
       hidden: $D^{(3)}\!\setminus\!D^{(1)}$}
    (D3);
\end{tikzpicture}
\caption{Relationships among the three dependency layers.
$D^{(1)}$ is extracted from the initial manifest, $D^{(2)}$ from the
final successfully installed environment, and $D^{(3)}$ from independent
runtime tracing via \texttt{strace}.
No fixed set-inclusion is assumed between $D^{(1)}$ and $D^{(2)}$.
Pairwise gaps define \emph{phantom} dependencies (declared but not needed),
\emph{hidden} dependencies (needed but not declared or installed), and
\emph{bloat} (installed but unused).}
\label{fig:layers}
\end{figure}
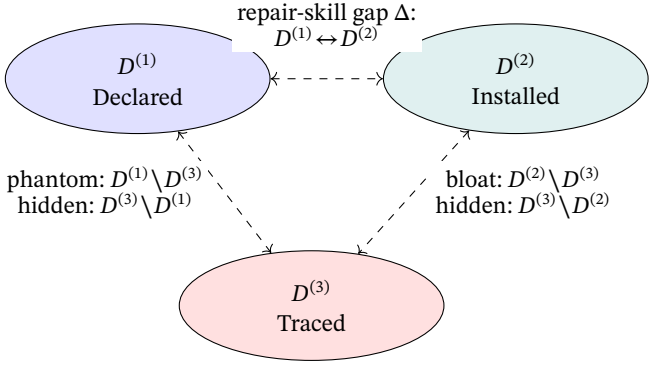

\begin{figure*}[t]
\centering
\includegraphics[width=0.75\textwidth]{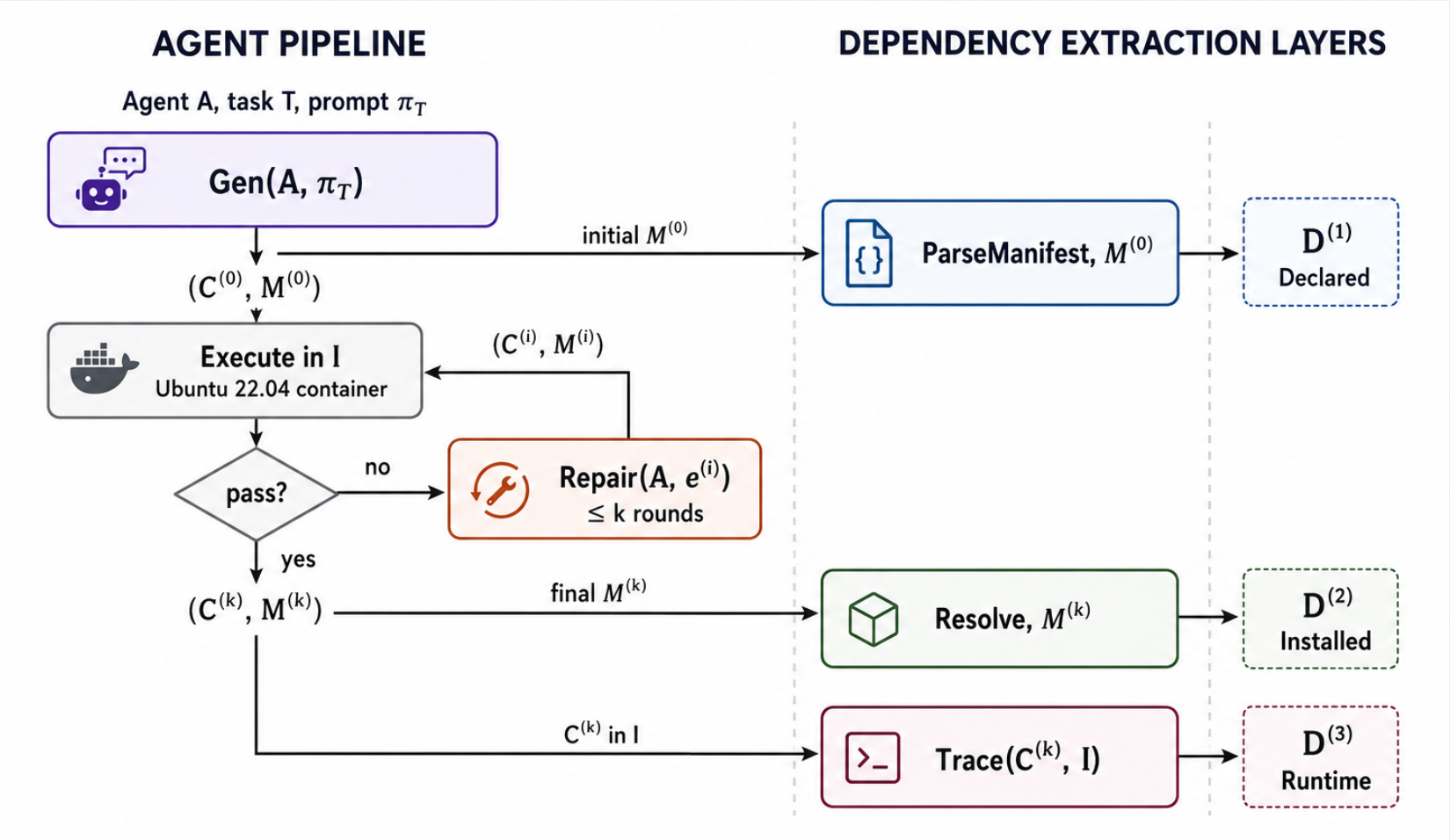}
\caption{Environment Evaluation Pipeline.  \textbf{Left:} the agent $A$ exercises its \texttt{Gen} skill to produce an initial project $(C^{(0)}, M^{(0)})$, which is executed in an isolated environment $I$. On failure, the \texttt{Repair} skill is invoked with the runtime error $e^{(i)}$ for up to $k$ rounds, yielding the final artifact $(C^{(k)}, M^{(k)})$. \textbf{Right:} three extraction primitives operate on the pipeline outputs to produce the three dependency layers: $D^{(1)}$ via manifest parsing of the initial declaration $M^{(0)}$; $D^{(2)}$ via package-manager resolution of the final manifest $M^{(k)}$; and $D^{(3)}$ via \texttt{strace}-based runtime tracing of $C^{(k)}$ in $I$, serving as ground truth.}
\label{fig:pipeline}
\end{figure*}

These three sets $D^{(1)}$,  $D^{(2)}$, and $D^{(3)}$ define our three-layer framework. To use these as a basis for measuring the generative and repair skills of coding agents, we do a pairwise comparison of $D^{(1)}$, $D^{(2)}$ and $D^{(3)}$. In this comparison, $D^{(3)}$ serves as a proxy for ground truth since obtaining $D^{(3)}$  is independent of the agent and this truth is obtained via the reference run of the executable corresponding to the code, which identifies dependencies for the specific execution run. 
%and is obtained via tooling that is independent of the specific language framework. We would like to emphasize that while we take $D^{3}$ as the ground truth, this truth is 

Comparing $D^{(1)}$ and $D^{(3)}$ in both directions reveals two complementary gaps: dependencies present in $D^{(1)}$ but absent from $D^{(3)}$ are \textit{phantom} dependencies \textit{i.e.,} declared by the agent's \texttt{Gen} skill but not actually needed by the code, while dependencies present in $D^{(3)}$ but absent from $D^{(1)}$ are \textit{hidden} dependencies \textit{i.e.,} necessary for execution but never declared. 

Comparing $D^{(2)}$ and $D^{(3)}$ exposes the agent's blindness to deeper dependencies: packages in $D^{(2)}$ but not $D^{(3)}$ constitute dependency \textit{bloat}. These are dependencies installed by the package manager but unused at runtime. Alternatively, packages in $D^{(3)}$ but not $D^{(2)}$ remain \textit{hidden} even after full resolution.

Finally, comparing $D^{(1)}$ and $D^{(2)}$ quantifies what the \texttt{Repair} skill contributes: packages added to or removed from the manifest across $k$ iterations, isolating the effect of iterative error correction from the agent's initial declaration. Figure~\ref{fig:layers} summarizes each pairwise comparison and the gap types each exposes. 

%Thus, comparing $D^{(1)}$ and $D^{(3)}$, we measure the declarative dependency gap between what the agent's \texttt{Gen} skill declares and what the code necessarily and sufficiently needs, also termed \textit{phantom} dependencies. By comparing $D^{(2)}$ and $D^{(3)}$ we measure the agent's lack of knowledge about deeper dependencies, such as transitive and implicit dependencies~\cite{x}, termed bloated dependencies. Finally, comparing $D^{(1)}$ and  $D^{(2)}$ provides a measure of the repair skill of the agents. Figure~\ref{} shows each pair of comparisons and the types of gaps within agent behavior understood by our pipeline. 

% Both of these comparisons are made in each direction, exposing different types of gaps. For example, 
% phantom dependencies~\cite{} appear in 
% $D^{(1)}$ but not $D^{(3)}$; hidden dependencies appear in $D^{(3)}$ but not in $D^{(1)}$; bloated 
%  dependencies~\cite{} appear in $D^{(2)}$ but not in $D^{(3)}$. 

%% file: sections/03_primitives.tex
\section{Protocol Primitives: Dependency Generation and Extraction}
\label{sec:primitives}

We now describe the different primitives used in our protocol, namely how \texttt{Gen} and \texttt{Repair} are configured with prompts and tasks, and how \texttt{ParseManifest}, \texttt{Resolve} and \texttt{Trace} lead to $D^{(1)}$,  $D^{(2)}$, and $D^{(3)}$. Figure~\ref{fig:pipeline} shows the use of these primitives as part of an agentic pipeline that includes \texttt{Gen} and \texttt{Repair}.

\subsection{Dependency Generation Primitives}
\label{sec:prompt}

The quality of the results from $\texttt{Gen}$ and $\texttt{Repair}$ depends on two inputs: the complexity of the programming task $T$ and the design of the prompt $\pi_T$. We address each through principled design criteria.

Tasks are drawn from common real-world programming problems, subject to two constraints: First, each task must be platform-independent i.e.,  implementable in every language under study  so that cross-language comparisons are 
controlled at the task level rather than confounded by structural differences between tasks. Second, each task must also be self-contained, requiring no external services, APIs, or credentials, so that dependency failures are attributable to the agent and not to environmental inaccessibility.

The prompt template instantiated for each task is shown in Figure~\ref{fig:prompt}. Every instance shares the same structure of task title, task description, and functional requirements. Further, there are two explicit framing instructions, differing only in the target language runtime and manifest filename, so that cross-agent comparisons are controlled at the prompt level. 

The two framing instructions specify that the agent must generate a \textit{complete} project for a \textit{clean} environment; the clean environment is language-specific and stated explicitly in every instance. Completeness is treated as an experimental variable instantiated in two prompt variations. In the \textit{implicit} variant, no completeness specification is given, measuring the agent's default behavior when a developer is succinct. In the \textit{explicit} variant, the agent is instructed to produce source code, a dependency manifest covering both direct and transitive dependencies pinned to exact versions, and a README — measuring behavior when a developer is meticulous. These two extremes reflect the observation that real-world developer prompts cluster toward one or the other rather than occupying a graded middle ground.

\begin{figure}[t]
\centering
\fbox{\parbox{0.93\columnwidth}{\normalsize\ttfamily\raggedright
\textbf{TASK:} \{task\_title\}\\[4pt]
\{task\_description\}\\[4pt]
\textbf{FUNCTIONAL REQUIREMENTS:}\\
\{functional\_requirements\}\\[4pt]
Create a complete \{language\} project for a clean Ubuntu 22.04
machine with only \{language\_runtime\} installed. Include:\\[2pt]
\hspace*{1em}--~Source code\\[1pt]
\hspace*{1em}--~\parbox[t]{\dimexpr0.93\columnwidth-2.5em}{\raggedright
    \{dependency\_file\} with all dependencies (direct and transitive)
    pinned to exact versions}\\[1pt]
\hspace*{1em}--~\parbox[t]{\dimexpr0.93\columnwidth-2.5em}{\raggedright
    README.md with setup instructions, dependency explanations, build
    steps, run commands, and expected output}
}}
\caption{A prompt template. The template is instantiated with language, manifest filename, and runtime; all functional
content is identical across languages.}
\label{fig:prompt}
\end{figure}

A prompt $\pi_T$ is considered valid if the agent produces code that passes an accompanying test suite specified within the functional requirements. This grounds prompt adequacy in an objective, reproducible criterion: passing tests confirm the prompt was sufficiently specified, ensuring that any dependency misspecification in the output is attributable to the agent rather than to an underspecified prompt.

\subsection{Dependency Extraction Primitives}

Amongst the three extraction primitives $\texttt{ParseManifest}_l$, $\texttt{Resolve}_l$ , and \texttt{Trace}, the first two are language-specific although we describe a generalized specification.  The \texttt{Trace} primitive is language-independent, but relies on the clean environment.   

%The three extraction primitives Parse\_l, Resolve\_l , and Trace extract dependency lists from manifests and code executables produced by the  Gen and Repair agent primitives and together produce the three dependency layers  $D^1$, $D^2$, and $D^3$.

\subsubsection{Parse Primitive}

A \textit{dependency manifest} is a language-ecosystem-specific
configuration file that formally declares the external packages a software
project requires to execute, and serves as the input artifact that package
managers read to install and resolve dependencies. In our framework, the
initial manifest $M^{(0)}$ is the agent's explicit machine-readable claim
of what the code needs in response to $\pi_T$, while the final manifest
$M^{(k)}$ is the manifest produced after iterative repair.

$\texttt{ParseManifest}_\ell$ comprises parsing rules and takes as input
$M^{(0)}$ to produce $D^{(1)}$, the declared dependency set. The parsing
rules are language-specific and currently support four language
ecosystems.
For Python, it reads each line of \texttt{requirements.txt} and strips
version constraints to normalize package names for cross-agent comparison.
For Java, it extracts every \texttt{<artifactId>} from the
\texttt{<dependencies>} section of \texttt{pom.xml}.
For JavaScript, it reads the keys of the \texttt{"dependencies"} object in
\texttt{package.json}, deliberately excluding \texttt{"devDependencies"}
since build-time tools are outside the runtime supply chain; this exclusion
is applied consistently across all agents and tasks.
For C++, it collects every library name from
\texttt{FetchContent\_Declare} and \texttt{find\_package} calls in
\texttt{CMakeLists.txt}. Note that C++ differs architecturally from the
other three languages in that CMake does not have a canonical package
manager, which has implications for $\texttt{Resolve}_\ell$ discussed
below.

Each rule is a deterministic extraction from a single field of a formally
specified file format and involves no heuristic inference; the complete
parsing implementation is released in our artifact. Because the same
manifests are subsequently consumed by each ecosystem's package manager in
the $\texttt{Resolve}_\ell$ step, any systematic parsing error would
surface as a mismatch between $D^{(1)}$ and the resolver's reported
dependency set during execution.

\subsubsection{Resolve primitive}
The primitive $\texttt{Resolve}_\ell$ determines $D^{(2)}$, the set of installed dependencies, within the isolated environment $I$. The Docker instance that instantiates $I$ is built from a base OS image with the language runtime, i.e., compiler, interpreter, or virtual machine, and the language-specific package manager. No additional packages are pre-installed beyond this toolchain baseline.

Within $I$, the primitive $\texttt{Resolve}_\ell$ executes the final manifest $M^{(k)}$, causing the language-specific package manager to install both the dependencies explicitly declared in $M^{(k)}$ and any additional transitive dependencies they require. $\texttt{Resolve}_\ell$ then queries the package manager for the complete list of packages installed in $I$, %excluding the pre-existing toolchain baseline, 
and reports this as $D^{(2)}$:
$$D^{(2)} := \mathrm{pkg\_list}_{post}(I, M^{(k)}) $$ %\setminus \mathrm{pkg_list}{pre}(I)$$. 

The correctness of $D^{(2)}$ rests on the package manager itself serving
as the measurement instrument: the reported set is the ecosystem's own
authoritative enumeration of installed packages, queried inside $I$
immediately after installation, rather than an inference made by our
tooling. The clean toolchain baseline of $I$ is fixed and identical across
all runs, so the enumeration reflects only packages installed from
$M^{(k)}$. A failed or partial resolution is not silently absorbed: it
surfaces as an installation error in the protocol and is returned to
\texttt{Repair} as $e^{(i)}$.

Given the declared dependency set $D^{(1)}$ obtained from $\texttt{ParseManifest}_\ell$, $\texttt{Resolve}_\ell$ also measures the transitive overhead introduced by the package manager:
$$\Delta = D^{(2)} \setminus D^{(1)};  \hspace{25pt} \rho =  \frac{|D^{(2)}|}{|D^{(1)}|}$$
where $\Delta$ is the set of packages installed that the agent did not explicitly declare, and $\rho$ is the inflation ratio measuring the multiplicative factor by which the package manager expands the agent's declared dependency set. $\rho = 1$ indicates no transitive resolution; $\rho \gg 1$ indicates aggressive transitive expansion. 

For example, if $M^{(k)}$ declares only \texttt{pandas} and the build succeeds on $M^{(0)}$ such that $M^{(k)} = M^{(0)}$, then pip additionally installs \texttt{numpy}, \texttt{python-dateutil}, \texttt{pytz}, \texttt{tzdata}, and \texttt{six} as transitive dependencies, giving $|D^{(1)}| = 1$, $|D^{(2)}| = 6$, $|\Delta| = 5$, and $\rho = 6.0$. Java and JavaScript behave analogously and are described in Section~\ref{sec:setup}.
C++ is a special case. Since CMake has no canonical package manager, no transitive dependency resolution occurs at the package manager layer. Consequently, $\mathrm{Resolve}_\ell$ for C++ reduces to $D^{(2)} := D^{(1)}$, meaning $\Delta = \emptyset$ and $\rho = 1$ by construction rather than by agent behavior. The $D^{(1)}$-vs-$D^{(2)}$ comparison is therefore structurally uninformative for C++ and is excluded from cross-language analyses of transitive dependency gaps.

\subsubsection{Trace primitive}

The $\texttt{Trace}$ primitive operates on the final successfully installed program $(C^{(k)}, M^{(k)})$ within the isolated environment $I$, determining which packages are actually loaded during a specific execution run. 
Unlike $\texttt{ParseManifest}_\ell$ and $\texttt{Resolve}_\ell$, the $\texttt{Trace}$ method is independent of both the language runtime and the coding agent: It is independent of the language-runtime due to the choice of the tracing method; It is independent of the coding agent as it does not query any agent-produced artifact but instead observes program behavior directly for used dependencies. 
It does however require that $\texttt{Resolve}_\ell$ has succeeded, since a successfully installed program is a precondition for tracing.

$\texttt{Trace}$ produces $D^{(3)}$, the set of dependencies necessary and sufficient for the observed execution of $C^{(k)}$ in $I$. Comparing $D^{(3)}$ against $D^{(1)}$ and $D^{(2)}$ exposes overclaimed dependencies i.e., packages declared or installed but never loaded at runtime. Formally:
\begin{equation}
\begin{split}
D^{(1)} \setminus D^{(3)} \quad \text{(declared but unused)};  \\ 
 D^{(2)} \setminus D^{(3)} \quad \text{(installed but unused)}
\end{split}
\end{equation}
While modern interpreted language IDEs often warn developers of unused declared dependencies, the problem is equally present in compiled languages such as Java and C++, where a package may be linked but never exercised on a given execution path~\citep{quach2018debloating}.

$\texttt{Trace}$ uses \texttt{ptrace} to intercept all file- and process-related system calls made by the running program and any child processes it spawns. Each dependency file accessed during execution is recorded in a log. 
Dependency-specific files are then filtered using a dictionary of language-specific installation paths, such as \texttt{dist-packages} for Python and \texttt{dependency} for Java  to map accessed file paths to package names, enabling direct comparison with $D^{(1)}$ and $D^{(2)}$. Language-specific path-to-package mapping rules are described in Section~\ref{sec:setup}.

\begin{center}
\begin{lstlisting}[language=C++,
  basicstyle=\fontsize{8.5}{10.5}\selectfont\ttfamily,
  tabsize=2, breaklines=true, breakatwhitespace=true,
  xleftmargin=0pt,
  caption={Example code showing reference execution}, label=Listing]
int main(int argc, char* argv[]) {
  if (std::string(argv[1]) == "--matrix") {
    // uses Eigen
    Eigen::MatrixXd m = Eigen::MatrixXd::Random(3, 3);
  } else if (std::string(argv[1]) == "--json") {
    // uses nlohmann/json
    nlohmann::json j = nlohmann::json::parse("{}");
  }
}
\end{lstlisting}
\end{center}

Two limitations apply to $D^{(3)}$. First, since tracing captures only dependencies loaded during a specific execution run, conditionally imported packages not exercised on the observed execution path are absent from $D^{(3)}$. See example in Listing~\ref{Listing}. We assume such conditional dependencies are rare in practice and do not materially affect aggregate agent behavior across tasks. 
Second, $\texttt{Trace}$ operates at the file access level and therefore cannot observe header-only libraries, which are included at compile time rather than loaded at runtime. 
Such libraries may appear in $D^{(1)}$ or $D^{(2)}$ but will be systematically absent from $D^{(3)}$, which is particularly relevant for C++ where header-only libraries are common. We acknowledge this as a scope limitation of dynamic over static analysis.

%% file: sections/04_setup.tex
\section{Experiment Setup}
\label{sec:setup}

In this section, we first instantiate the protocol of Section~\ref{sec:protocol} with tasks, agents, and language ecosystems. We then describe how the agent protocol is conducted via configured primitives of Section~\ref{sec:primitives}. The experimental setup of tasks and extracted dependencies will be made available as an open reproducible artifact.

\subsection{Protocol Instantiation and Execution}

\vspace{5pt}
\textbf{Programming Tasks $T$.} We instantiate the task set ${T}$ with 50 benchmark programming tasks spanning eight application domains, as shown in Table~\ref{tab:domains}. Tasks are selected to satisfy the two criteria established in Section~\ref{sec:prompt}: platform independence, ensuring each task can be implemented identically across all four language ecosystems in $\mathcal{L}$, and self-containment, requiring no external services, APIs, or credentials so that dependency misspecification is attributable to the agent's \texttt{Gen} and \texttt{Repair} skills rather than environmental inaccessibility. The eight domains are chosen to provide broad coverage of real-world software categories that developers routinely delegate to coding agents, ranging from data processing and cryptography to system utilities and scientific computing. The distribution across domains, which is weighted toward data processing and cryptography reflects the relative frequency of such tasks in open-source coding benchmarks~\cite{chen2021codex, austin2021mbpp}. The 50-task scale is chosen to ensure statistical power for cross-agent and cross-language comparisons while remaining feasible for the fully manual execution protocol described below.

\begin{table}[t]
\caption{The 50 benchmark programming tasks span eight application domains.}
\label{tab:domains}
\centering
\small
\setlength{\tabcolsep}{4pt}
\begin{tabular}{@{}lcp{5.0cm}@{}}
\toprule
\rowcolor{HeadTint}
\textbf{Domain} & \textbf{$n$} & \textbf{Representative tasks} \\
\midrule
Data Processing  & 10 & CSV statistical analyzer, JSON schema validator,
                        XML-to-JSON converter, YAML config merger \\
Cryptography     &  7 & AES-256 file encryption, RSA digital signatures,
                        TOTP generator, bcrypt hasher \\
Image Processing &  6 & Format converter, watermarking, histogram analyzer,
                        EXIF metadata extractor \\
Networking       &  6 & HTTP downloader with retry, port scanner,
                        DNS resolver, static file server \\
Text Processing  &  6 & Markdown-to-HTML, spell checker, TF-IDF search
                        engine, template engine \\
Compression      &  5 & ZIP archive manager, GZIP tool, file deduplicator,
                        compression benchmark \\
Math/Scientific  &  5 & Matrix operations, linear regression,
                        Dijkstra shortest path, FFT analyzer \\
System Utilities &  5 & QR code generator/reader, PDF report builder,
                        Merkle tree verifier, cron parser \\
\bottomrule
\end{tabular}
\end{table}

\textbf{Language Ecosystems $\mathcal{L}$}. The language ecosystem set consists of $\mathcal{L} = \{\text{Python}, \text{Java}, \text{JavaScript}, \text{C++}\}$. These language ecosystems are used for configurations of $\texttt{ParseManifest}_\ell$, $\texttt{Resolve}_\ell$, and $\texttt{Trace}$ defined in Section~\ref{sec:primitives}. These four ecosystems were selected to represent diversity in runtimes (interpreted vs compiled), package manager design (centralized vs decentralized), and transitive resolution behavior ($\rho \gg 1$ for npm vs $\rho \approx 1$ for CMake), enabling cross-ecosystem analysis of the environment specification gap.

\vspace{5pt}
\textbf{Coding agents $\mathcal{A}$.} The agent set is $\mathcal{A} = \{a_1, a_2, a_3\}$, instantiated as:

\begin{itemize}[nosep,leftmargin=*]
  \item $a_1$: \textbf{Claude Code} (Anthropic, Claude Opus with extended thinking).
  \item $a_2$: \textbf{Codex} (OpenAI, the latest Codex model via GitHub Copilot).
  \item $a_3$: \textbf{Gemini Code Assist} (Google, Gemini with the highest reasoning configuration).
\end{itemize}

These three agents are selected representing the current frontier of coding agent capability, covering the three dominant providers, and are accessed exclusively through their command-line interfaces at default settings. This mirrors the experience of a developer using these agents in practice. All agents receive the same prompt $\pi_T$ for each task, with the iteration budget $k = 10$ repair cycles specified in each agent's configuration file as described in Section~\ref{sec:protocol}, yielding a maximum of 11 agent interactions per instance — one $\texttt{Gen}$ call followed by up to 10 $\texttt{Repair}$ calls. Further, agents preserve full history within the context window during the \texttt{Repair} cycle, enabling the agent to reason about the sequence of previously attempted repairs. 

%At each failed iteration $i$, the verbatim terminal error $e^{(i-1)}$ is appended to the ongoing agent conversation with full history preserved, enabling the agent to reason about the sequence of previously attempted repairs. The accumulated conversation history across up to $k = 10$ repair iterations remains within the context window of all three agents \textbf{[verify and cite or qualify]}. The loop terminates at the first $i \leq k$ for which $e^{(i)} = \emptyset$, defining $(C^{(k)}, M^{(k)})$ as in Section~\ref{sec:protocol}. If no such $i$ exists within budget, the instance is marked failed and excluded from $D^{(2)}$ and $D^{(3)}$ analyses; $D^{(1)}$ remains defined for all instances regardless of execution outcome, since it is derived from $M^{(0)}$ prior to any execution attempt. This asymmetry implies that cross-agent comparisons of $D^{(2)}$ and $D^{(3)}$ are computed only over the subset of instances where all agents succeed, to ensure comparability.

\vspace{5pt}
\textbf{Trial Structure and Study Scale.} Every agent $a \in \mathcal{A}$ receives every prompt $\pi_T$ for every task-language pair $(T, \ell) \in \mathcal{T} \times \mathcal{L}$, yielding $|\mathcal{T}| \times |\mathcal{L}| \times |\mathcal{A}| = 50 \times 4 \times 3 = 600$ primary instances. To assess the stochastic variability of the \texttt{Gen} and \texttt{Repair} skills, $a_1$ is additionally executed for two further independent trials per task-language pair, each initialized from a fresh conversation with no memory of prior attempts, contributing $50 \times 4 \times 2 = 400$ stochastic trials. $a_1$ is selected for variance analysis as it is the only agent achieving 100\% final execution success (Section~\ref{sec:results}), guaranteeing complete data across all task-language pairs for a statistically meaningful variance analysis. The total study scale is thus 1000 evaluation runs (Table~\ref{tab:scale}).

Section~\ref{sec:prompt} defines two prompt variations: \textit{implicit}, in which no completeness specification is provided, and \textit{explicit}, in which the agent is instructed to produce source code, a pinned dependency manifest, and a README. All 600 primary instances use the explicit variant, so the reported results characterize agent behavior under maximal prompt specificity; this constitutes the agent's best case, and the misspecification rates reported in Section~\ref{sec:results} should therefore be read as lower bounds on what less specific prompting would produce. A systematic comparison against the implicit variant is left to future work.

\begin{table}[!htb]
\caption{Study scale.}
\label{tab:scale}
\centering
\small
\begin{tabular}{lr}
\toprule
\rowcolor{HeadTint}
\textbf{Dimension} & \textbf{Count} \\
\midrule
Benchmark programming tasks                              &  50 \\
Target language ecosystems                               &   4 \\
LLM coding agents                                        &   3 \\
Primary configurations ($50{\times}4{\times}3$)          & 600 \\
Stochastic trials (Claude Code, 2 extra ${\times}$ 200)  & 400 \\
Maximum agent interactions per instance                  &  11 \\
\midrule
\rowcolor{SumTint}
\textbf{Total evaluation runs}                           & \textbf{1000} \\
\bottomrule
\end{tabular}
\end{table}

\vspace{5pt}
\textbf{Isolated Environment $I$.} The isolated environment $I$ is
realized as a per-language Docker container built from the
\texttt{ubuntu:22.04} base image and containing only the language
toolchain necessary to execute $\texttt{Resolve}_\ell$ and
$\texttt{Trace}_\ell$:
\begin{itemize}[nosep,leftmargin=*]
\item ${I}_{\mathit{Python}}$ (\texttt{tmlr-python-eval}): Python 3.10, pip
\item ${I}_{\mathit{Java}}$ (\texttt{tmlr-java-eval}): JDK 17, Maven 3
\item ${I}_{\mathit{JS}}$ (\texttt{tmlr-js-eval}): Node.js 20 LTS, npm
\item ${I}_{\mathit{C++}}$ (\texttt{tmlr-cpp-eval-v2}): G++ 12, CMake 3.22,
      \texttt{make}, \texttt{git}, \texttt{zlib1g-dev}, and standard headers
\end{itemize}
No image contains common system development libraries beyond those
listed: packages such as \texttt{OpenSSL}, \texttt{libcurl}, or
\texttt{SQLite3} development headers are absent from every container, so
any agent assumption of their presence surfaces as an observable failure
rather than silently succeeding. The C++ container necessarily includes
\texttt{zlib1g-dev} and standard C/C++ headers as these are required by
the CMake toolchain; these are recorded in $\mathrm{pkg\_list}_{pre}(I)$
and subtracted from $D^{(2)}$ via the baseline procedure defined in
Section~\ref{sec:primitives}, ensuring they do not contaminate the
installed dependency measurement. Each image additionally contains
Sciunit~\citep{pham2013using}, whose \texttt{ptrace}-based engine
implements the $\mathrm{Trace}_\ell$ primitive inside the same container;
Sciunit is part of the fixed baseline, and its own file accesses are
excluded from $D^{(3)}$ so that the runtime trace reflects only the
agent-generated artifact.

\vspace{5pt}
\textbf{Manual execution.}
All 1000 instances were conducted manually, with each instance comprising a multi-turn interaction: the initial \texttt{Gen} prompt followed by up to ten \texttt{Repair} turns, for a maximum of eleven agent interactions per instance (Table~\ref{tab:scale}). This means the following actions were undertaken by hand: (i) the initial prompt and every subsequent repair turn were submitted to each agent's CLI, (ii) generated artifacts were transferred into the Docker containers hosted on an AWS EC2 instance, (iii) runtime errors were transcribed verbatim into the ongoing agent conversation, and (iv) outcomes were recorded into structured CSV files. Automation was deliberately avoided since in our experience the diversity of build failures across four ecosystems was too broad for reliable scripting, and an automated harness would either mask real failure modes through generic recovery heuristics or accumulate so many special-case rules that the harness itself would become a confounding factor, especially in the evaluation of the \texttt{Repair} skill. Once the manual execution of a project is completed and the necessary dependencies collected, each container is destroyed via \texttt{docker run -{}-rm} and a fresh container is instantiated for the next programming task.

\subsection{Primitive Configuration}

\vspace{5pt}
\noindent \textbf{$\texttt{ParseManifest}_l$ and $\texttt{Resolve}_l$}. Table~\ref{tab:extraction} shows concrete instantiations of $\texttt{ParseManifest}_l$ and $\texttt{Resolve}_l$ per ecosystem.

\vspace{5pt}
\noindent
\textbf{\texttt{Trace}.} We realize $\texttt{Trace}(C^{(k)}, I)$ using Sciunit~\citep{pham2013using}, compiled from source and installed in every Docker image. Sciunit attaches to the target process via the Linux \texttt{ptrace} API and intercepts all \texttt{open()}, \texttt{openat()}, \texttt{read()}, and \texttt{execve()} system calls, writing the absolute path of each accessed file to a provenance log. \texttt{READ}-tagged entries are filtered using ecosystem-specific path rules to map accessed file paths to package names, enabling direct comparison with $D^{(1)}$ and $D^{(2)}$.

A critical implementation challenge arises for Java. LLM-generated Java projects frequently use the Maven Shade Plugin to produce a fat JAR bundling all dependency classes into a single archive. Under fat JAR invocation, the JVM issues exactly one \texttt{openat()} call for the archive itself, making the provenance log opaque to dependency analysis. Sciunit cannot determine which packages were loaded since all are bundled inside the same file. We refer to this as the \emph{fat JAR opacity problem}. To resolve it, we replace fat JAR execution with classpath-based execution before tracing: dependency JARs are unpacked as physically separate files via \texttt{mvn dependency:copy-dependencies}, and the program is traced against the unpacked classpath. With each JAR a distinct file, the JVM issues a separate \texttt{openat()} per loaded JAR, which appears individually in the provenance log. We verified that classpath-based execution produces functionally identical output to fat JAR execution in all traced cases.

For \textbf{C++}, the compiled binary is traced directly. At startup, the dynamic linker issues \texttt{openat()} calls for every shared library the binary was linked against. We retain \texttt{.so} entries and exclude the seven base system libraries loaded by every C++ binary regardless of application code: \texttt{libc.so}, \texttt{libstdc++.so}, \texttt{libm.so}, \texttt{libgcc\_s.so}, \texttt{libpthread.so}, \texttt{libdl.so}, and \texttt{librt.so}, leaving only application-level libraries in $D^{(3)}$. The header-only library limitation noted in Section~\ref{sec:primitives} is most acute here: \texttt{nlohmann\_json}, the most frequently declared C++ dependency across all agents (65 successful manifests), is compiled directly into the binary and produces no \texttt{.so} file, and therefore no provenance log entry. For projects whose declared dependencies are entirely header-only, $D^{(3)} = \emptyset$ even when $D^{(1)}$ is non-empty. We return to this limitation in Section~\ref{sec:limitations}.

\begin{table*}[t]
\caption{Per-ecosystem realization of $\mathrm{ParseManifest}_\ell$,
$\mathrm{Resolve}_\ell$, and $\mathrm{Trace}_\ell$ for the three
dependency layers.}
\label{tab:extraction}
\centering
\footnotesize
\setlength{\tabcolsep}{5pt}
\begin{tabularx}{\textwidth}{@{}l
  >{\raggedright\arraybackslash}X
  >{\raggedright\arraybackslash}X
  >{\raggedright\arraybackslash}X@{}}
\toprule
\rowcolor{HeadTint}
\textbf{Ecosystem}
  & \textbf{Layer~1: $D^{(1)} = \mathrm{ParseManifest}_\ell(M^{(0)})$}
  & \textbf{Layer~2: $D^{(2)} = \mathrm{Resolve}_\ell(M^{(k)})$}
  & \textbf{Layer~3: $D^{(3)} = \mathrm{Trace}_\ell(C^{(k)}, I)$} \\
\midrule
\textcolor{LPy}{\textbf{Python}}
  & Parse \texttt{requirements.txt} line by line; strip version
    constraints
  & Diff of \texttt{pip list} before vs.\ after \texttt{pip install}
  & Syscall \texttt{READ}s under \texttt{dist-packages/} or
    \texttt{site-packages/} \\[5pt]
\textcolor{LJava}{\textbf{Java}}
  & \texttt{<artifactId>} values in \texttt{pom.xml}
    \texttt{<dependencies>}
  & \texttt{mvn dependency:list} artifact IDs (full transitive closure)
  & Syscall \texttt{READ}s of \texttt{target/dependency/*.jar} via
    classpath execution \\[5pt]
\textcolor{LJS}{\textbf{JavaScript}}
  & Keys of \texttt{"dependencies"} in \texttt{package.json}
  & \texttt{npm list -{}-all -{}-json}; full \texttt{node\_modules/} tree
  & Syscall \texttt{READ}s under \texttt{node\_modules/} \\[5pt]
\textcolor{LCpp}{\textbf{C++}}
  & \texttt{FetchContent\_Declare} and \texttt{find\_package} names in
    \texttt{CMakeLists.txt}
  & CMake configure log
    ($D^{(2)} \approx D^{(1)}$; transitive expansion typically absent)
  & Syscall \texttt{READ}s of \texttt{.so} files; seven base system
    libraries excluded \\
\bottomrule
\end{tabularx}
\end{table*}

%% file: sections/05_metrics.tex
\section{Evaluation Dimensions}
\label{sec:metrics}
% ═══════════════════════════════════════════════════════════════════════
We evaluate the agent evaluation protocol and the extraction primitives across five dimensions:
\begin{itemize}[leftmargin=*]
\item \textbf{Protocol Convergence.} The \texttt{Gen} $\rightarrow$ \texttt{Repair} cycle produces a successful artifact $(C^{(k)}, M^{(k)})$ through iterative resolution. We evaluate whether and how quickly an agent converges, i.e., reaches $e^{(k)} = \emptyset$ within budget, and how convergence rate varies across agents, languages, and task domains. This includes measuring both the binary outcome indicating if the cycle converges, and the number of iterations consumed via repair depth $i^* = \min\{i \leq k : e^{(i)} = \emptyset\}$.

\item \textbf{Environment Specification Quality.} The three dependency sets $D^{(1)}$, $D^{(2)}$, and $D^{(3)}$ produced by the extraction primitives form the basis for measuring specification quality. We evaluate two complementary measures. First, pairwise comparison of $\{D^{(1)}, D^{(2)}, D^{(3)}\}$ against $D^{(3)}$ as ground truth, quantifying the extent of phantom ($D^{(1)} \setminus D^{(3)}$), hidden ($D^{(3)} \setminus D^{(1)}$), bloated ($D^{(2)} \setminus D^{(3)}$), and transitive ($\Delta = D^{(2)} \setminus D^{(1)}$) dependencies, reported as precision, recall, and F1 scores. Second, the direction and magnitude of misspecification i.e., over- versus under-specification, measured via $\Delta$ and $\rho$. We additionally assess stochastic variability within a single agent using the repeated trials, determining whether misspecification is a deterministic property of the agent's knowledge or a stochastic artifact of generation.
measured via $\Delta$ and $\rho$. 
We additionally assess stochastic variability within a single agent using the repeated trials
%of $a_1$
, 
determining whether misspecification is a deterministic property of the agent's knowledge or a stochastic artifact of generation.

\item \textbf{Cross-Agent Comparison.} Given per-agent specification quality, we evaluate how agents differ from one another for identical task-language pairs. This covers pairwise dependency set agreement across agents and whether differences are systematic or task-dependent, isolating agent-specific behavior from task and language effects.
\item \textbf{Cross-Language Variation.} We evaluate whether specification quality varies systematically across the four language ecosystems $\mathcal{L} = \{\text{Python, Java, JavaScript, C++}\}$. If agents perform poorly and that poor performance is language-specific, this evaluation informs whether agents require language-specific training or whether general environment specification ability is the primary deficiency.
\item \textbf{Task Complexity Sensitivity.} We evaluate whether specification quality degrades with task complexity or varies by programming domain. This determines whether agents fail consistently on particular task types, providing diagnostic signal about which categories of software projects are most vulnerable to environment specification failure.
\end{itemize}

%% file: sections/06_results.tex
% ═══════════════════════════════════════════════════════════════════════
\section{Evaluation Results}
\label{sec:results}
% ═══════════════════════════════════════════════════════════════════════
We present our results across the five evaluation dimensions of
Section~\ref{sec:metrics}. The building blocks are already in place: the
dependency sets $D^{(1)}$, $D^{(2)}$, $D^{(3)}$ come from
Section~\ref{sec:protocol}, and the inflation ratio $\rho$ from
Section~\ref{sec:primitives}. Each subsection below only adds the small
amount of extra notation it needs.

We fix three conventions used throughout. First, subscripts identify the
setting: $D^{(j)}_{a,p,\ell}$ is the dependency set of agent $a$ on task
$p$ in language $\ell$. Second, a project counts as \emph{successful} when
its final artifact $(C^{(k)}, M^{(k)})$ runs without error in the clean
environment $\mathcal{D}_\ell$ within the repair budget; we write
$\mathrm{Success}(C, M, \mathcal{D}_\ell)$ for this condition and
$\mathcal{S}_{a,\ell}$ for the set of tasks where agent $a$ succeeds in
language $\ell$. Third, unless stated otherwise, every reported number is
a mean over successful projects.

% ─────────────────────────────────────────────────────────────────────
\subsection{Protocol Convergence}
\label{sec:conv}
% ─────────────────────────────────────────────────────────────────────

\begin{figure*}[t]
\centering
\includegraphics[width=\textwidth]{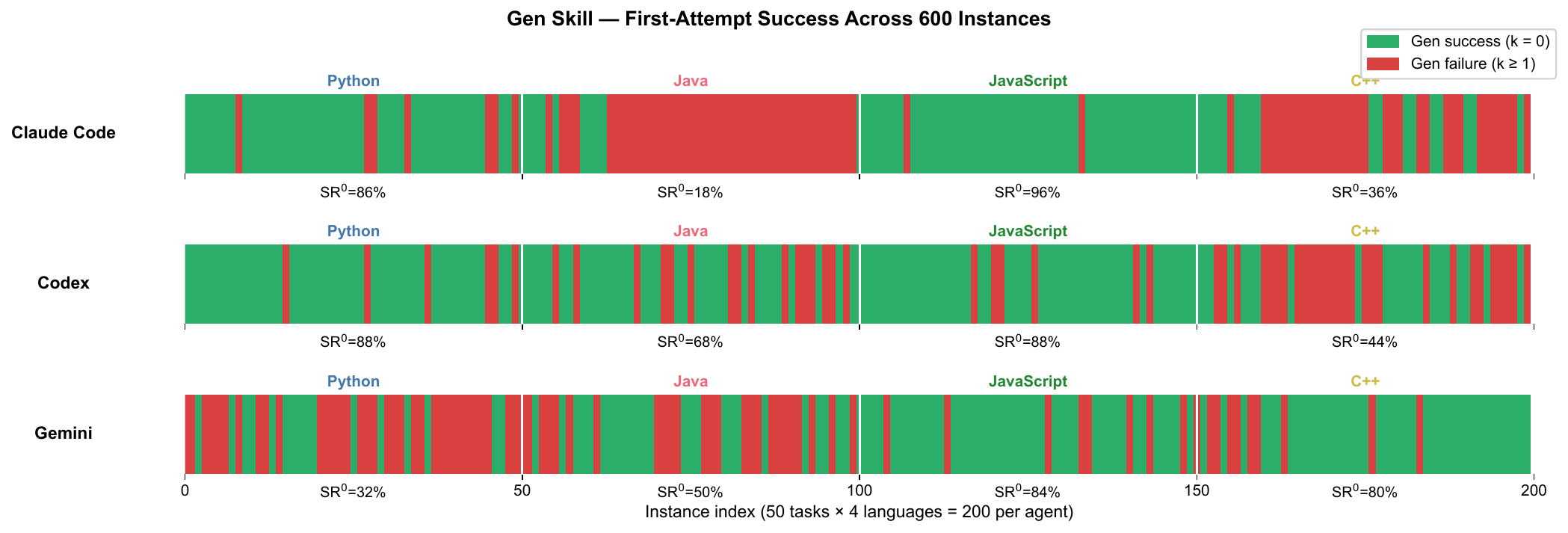}
\caption{\texttt{Gen} skill: first-attempt success across all 600 instances
(3 agents $\times$ 4 languages $\times$ 50 tasks). Each cell is one (agent,
language, task) triple; green indicates first-attempt success and red
indicates failure.}
\label{fig:gen_skill}
\end{figure*}

\begin{figure*}[t]
\centering
\includegraphics[width=\textwidth]{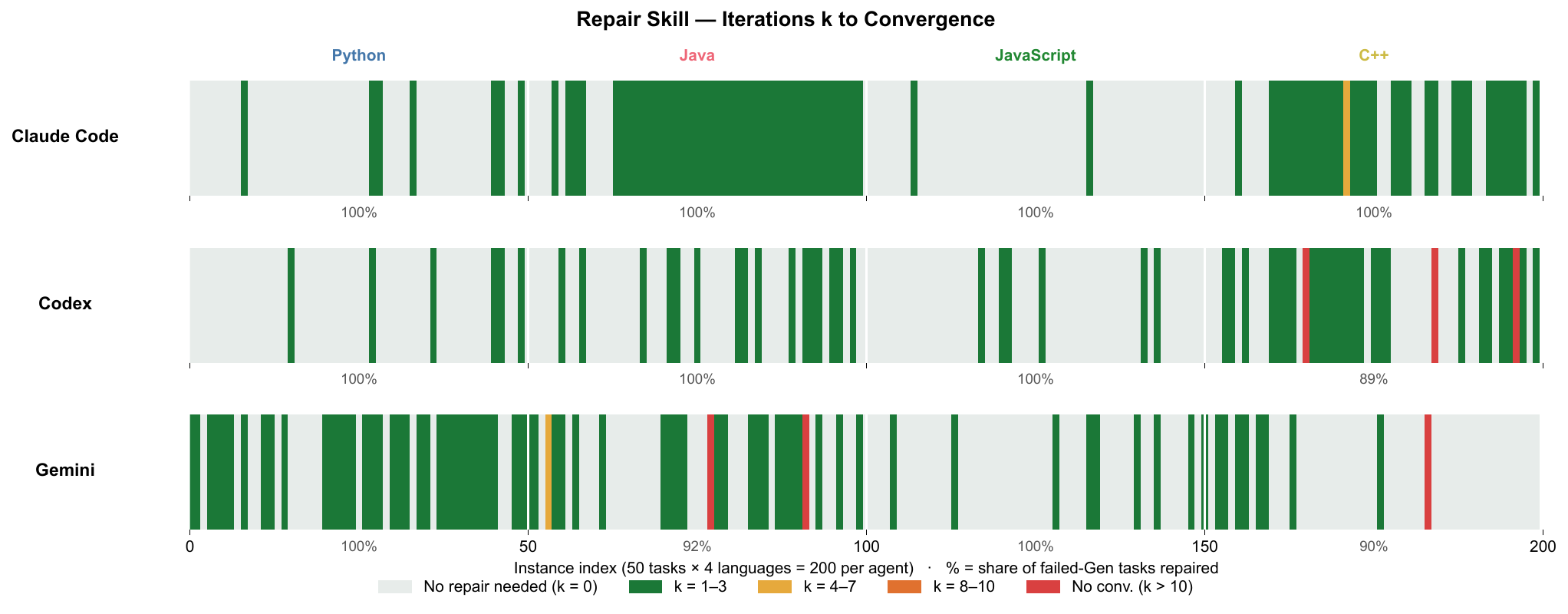}
\caption{\texttt{Repair} skill: distribution of repair cycles needed to
converge for each (agent, language) pair, bucketed into first-attempt
success, 1--3, 4--7, and 8--10 cycles, and failure within the budget.}
\label{fig:repair_skill}
\end{figure*}

\begin{figure}[t]
\centering
\includegraphics[width=\columnwidth]{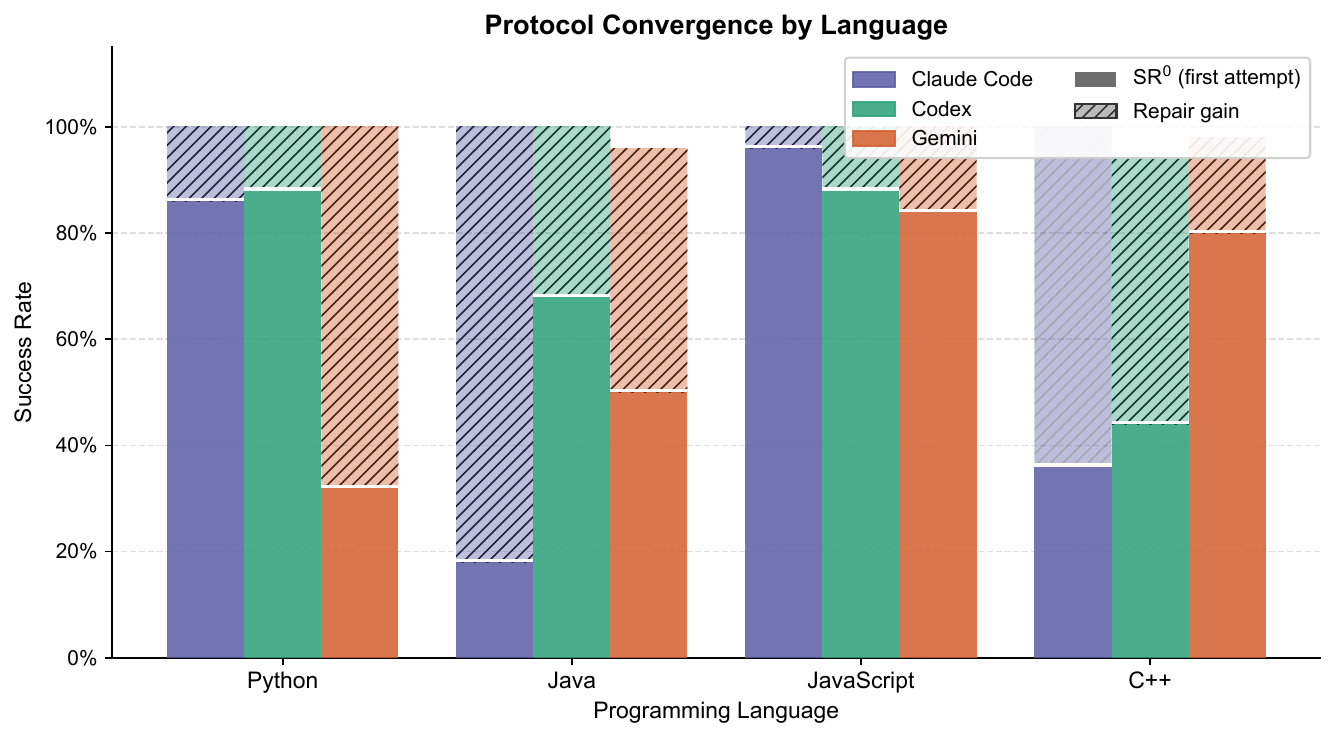}
\caption{Protocol convergence by language. Solid bars show the
first-attempt success rate $\mathrm{SR}^{(0)}$ and hatched extensions show
the self-correction gain contributed by \texttt{Repair}.}
\label{fig:conv_language}
\end{figure}

Figures~\ref{fig:gen_skill}--\ref{fig:conv_language} examine the agents for
protocol convergence. We first measure the performance of the \texttt{Gen}
skill, \textit{i.e.,} obtaining a successful software artifact for a
programming task in the first attempt. We measure this across the 600
instances, ordered first by agent and then by language, so that each cell in
Figure~\ref{fig:gen_skill} represents one (agent, language, task) triple.
Dense green regions show that Claude and Codex have a strong \texttt{Gen}
skill in Python and JavaScript but a much weaker one in Java and C++,
while Gemini inverts the pattern: it is weakest in Python (32\%
first-attempt success) yet the strongest of the three agents in C++
(80\%). No agent is strong in all four languages, and the failure patterns
differ across agents, motivating a closer examination of the
\texttt{Repair} skill.

Figure~\ref{fig:repair_skill} shows how much of the \texttt{Repair} skill
is exercised per agent. For each (agent, language) pair, it reports the
fraction of tasks resolved on the first attempt and the fraction requiring
$k$ repair cycles, with $k$ bucketed into the ranges $[1,3]$, $[4,7]$, and
$[8,10]$; the final category marks tasks that failed within the repair
budget. The vast majority of repairs complete within 1--3 cycles in every
ecosystem. Deep repair chains of four or more cycles are rare and appear
almost exclusively in C++. Claude repairs every one of its failed
first attempts in all four languages, whereas Codex recovers 89\% of its
failed C++ attempts and Gemini recovers 92\% in Java and 90\% in C++;
outright failures are therefore confined to Codex and Gemini in C++ and to
Gemini in Java.

Figure~\ref{fig:conv_language} aggregates these results by language. To
aggregate, we define a success rate at iteration $i$ for agent $a$ and
language $\ell$ as
\begin{equation}
  \mathrm{SR}^{(i)}_{a,\ell}
  \;=\;
  \frac{1}{|\mathcal{P}|}
  \sum_{p\,\in\,\mathcal{P}}
  \mathbf{1}\!\left[
    \mathrm{Success}\!\left(
      C^{(i)}_{a,p,\ell},\, M^{(i)}_{a,p,\ell},\, \mathcal{D}_\ell
    \right)
  \right],
  \label{eq:sri}
\end{equation}
so that $\mathrm{SR}^{(0)}_{a,\ell}$ is the first-attempt success rate,
measuring the \texttt{Gen} skill alone, and
$\mathrm{SR}^{(\leq k)}_{a,\ell} = \max_{i \leq k}
\mathrm{SR}^{(i)}_{a,\ell}$ is the final success rate after up to $k$
repair cycles. The self-correction gain
$\mathrm{SR}^{(\leq k)}_{a,\ell} - \mathrm{SR}^{(0)}_{a,\ell}$ measures the
contribution of the \texttt{Repair} skill to eventual convergence.

Figure~\ref{fig:conv_language} shows solid bars indicating
$\mathrm{SR}^{(0)}$ and hatched bars indicating the self-correction gain
for each agent and language. First-attempt success varies widely: Claude
ranges from 96\% in JavaScript and 86\% in Python down to 36\% in C++ and
only 18\% in Java, Codex spans 44\% (C++) to 88\% (Python and JavaScript),
and Gemini runs from 32\% (Python) to 84\% (JavaScript). After repair,
nearly every configuration converges: Claude reaches 100\% in all four
ecosystems, Codex falls short only in C++ (94\%), and Gemini falls
slightly short in Java (96\%) and C++ (98\%). The \texttt{Repair} skill
thus rescues most first-attempt failures, but the size of the hatched
region shows how much iterative feedback each configuration consumes to get
there. The corresponding analysis by task domain is deferred to
Section~\ref{sec:complexity}.

These results carry a direct implication for how coding agents are
evaluated. A benchmark that scores only final success would rate all three
agents as near-perfect, yet the cost of reaching that success differs by up
to a factor of five in repair cycles, and every repair cycle is additional
latency, compute, and opportunity for the agent to mutate the dependency
set. First-attempt environment specification is therefore a hidden cost
driver that functional-correctness benchmarks do not surface, and the first
manifest, not the final one, determines how much of that cost a deployment
pays.

% ─────────────────────────────────────────────────────────────────────
\subsection{Environment Specification Quality}
\label{sec:quality}
% ─────────────────────────────────────────────────────────────────────

\begin{figure}[t]
\centering
\includegraphics[width=\columnwidth]{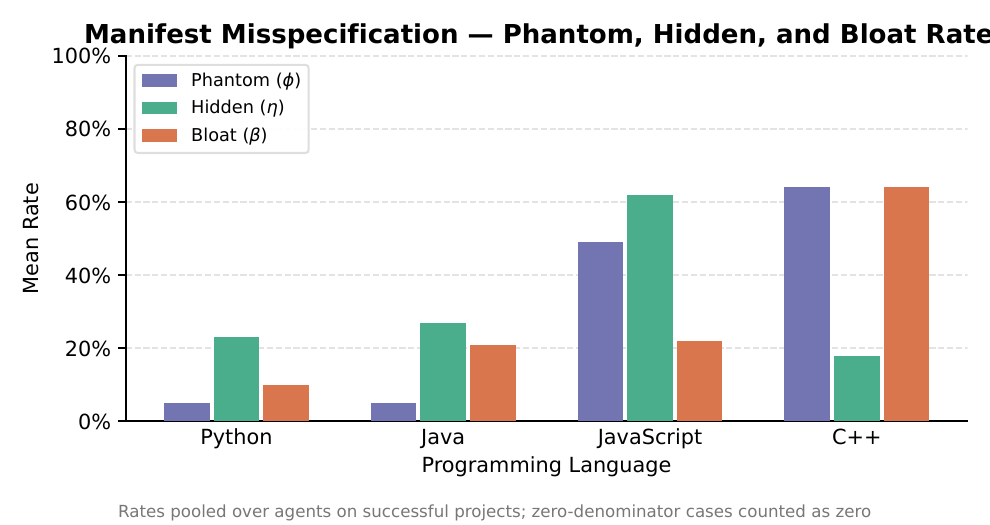}
\caption{Phantom, hidden, and installed-but-unused bloat rates by
ecosystem, averaged over successful projects with zero-denominator cases
included as zero.}
\label{fig:m3_phb}
\end{figure}

\begin{figure}[t]
\centering
\includegraphics[width=\columnwidth]{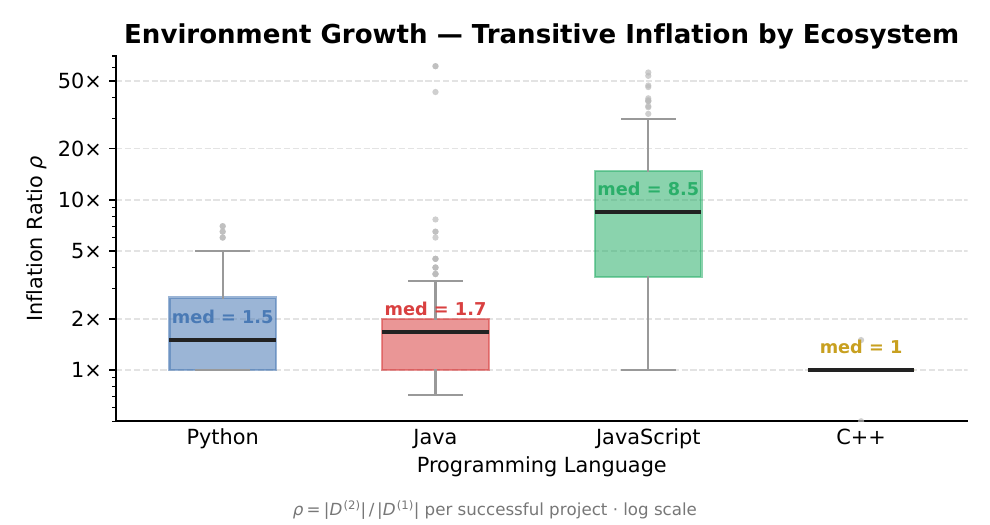}
\caption{Initial-to-final environment growth by ecosystem. Each box
summarizes the distribution of the inflation ratio $\rho$ of
Section~\ref{sec:primitives} over all successful projects in that
ecosystem (log scale).}
\label{fig:m2_inflation}
\end{figure}

\begin{table}[t]
\caption{Initial-manifest accuracy: treating runtime dependencies
$D^{(3)}$ as ground truth.}
\label{tab:manifest_accuracy}
\centering
\small
\begin{tabular}{llccc}
\toprule
\rowcolor{HeadTint}
\textbf{Agent} & \textbf{Language} & \textbf{Precision} & \textbf{Recall} & \textbf{F1} \\
\midrule
  \textcolor{AgClaude}{\textbf{Claude}} & \textcolor{LPy}{Python} & 0.990 & 0.642 & 0.733 \\
  \textcolor{AgClaude}{\textbf{Claude}} & \textcolor{LJava}{Java} & 0.935 & 0.739 & 0.758 \\
  \textcolor{AgClaude}{\textbf{Claude}} & \textcolor{LJS}{JavaScript} & 0.929 & 0.406 & 0.493 \\
  \textcolor{AgClaude}{\textbf{Claude}} & \textcolor{LCpp}{C++} & 0.037 & 0.781 & 0.031 \\
  \textcolor{AgCodex}{\textbf{Codex}} & \textcolor{LPy}{Python} & 0.938 & 0.827 & 0.837 \\
  \textcolor{AgCodex}{\textbf{Codex}} & \textcolor{LJava}{Java} & 0.939 & 0.684 & 0.735 \\
  \textcolor{AgCodex}{\textbf{Codex}} & \textcolor{LJS}{JavaScript} & 0.214 & 0.143 & 0.143 \\
  \textcolor{AgCodex}{\textbf{Codex}} & \textcolor{LCpp}{C++} & 0.062 & 0.658 & 0.004 \\
  \textcolor{AgGemini}{\textbf{Gemini}} & \textcolor{LPy}{Python} & 0.899 & 0.721 & 0.756 \\
  \textcolor{AgGemini}{\textbf{Gemini}} & \textcolor{LJava}{Java} & 0.947 & 0.523 & 0.624 \\
  \textcolor{AgGemini}{\textbf{Gemini}} & \textcolor{LJS}{JavaScript} & 0.040 & 0.182 & 0.005 \\
  \textcolor{AgGemini}{\textbf{Gemini}} & \textcolor{LCpp}{C++} & 0.059 & 0.708 & 0.004 \\
  \midrule
  \rowcolor{SumTint} \textbf{All} & \textcolor{LPy}{\textbf{Python}} & \textbf{0.944} & \textbf{0.713} & \textbf{0.765} \\
  \rowcolor{SumTint} \textbf{All} & \textcolor{LJava}{\textbf{Java}} & \textbf{0.940} & \textbf{0.653} & \textbf{0.708} \\
  \rowcolor{SumTint} \textbf{All} & \textcolor{LJS}{\textbf{JavaScript}} & \textbf{0.413} & \textbf{0.258} & \textbf{0.219} \\
  \rowcolor{SumTint} \textbf{All} & \textcolor{LCpp}{\textbf{C++}} & \textbf{0.049} & \textbf{0.729} & \textbf{0.018} \\
\bottomrule
\end{tabular}
\end{table}

All measures in ~\ref{tab:manifest_accuracy} are aggregated as means per stratum $(a, \ell)$ and per ecosystem $\ell$ pooled over agents. Because F1 is computed per project and then averaged, the tabulated F1 need not equal the harmonic mean of the tabulated precision and recall; the discrepancy is largest in C++, where many projects have zero declared--runtime overlap.

We measure environment specification quality in two complementary ways,
building directly on the three-layer framework of
Section~\ref{sec:protocol}. First, we operationalize the phantom, hidden,
and bloat categories of Section~\ref{sec:protocol} as rates, together with
the inflation ratio $\rho$ of Section~\ref{sec:primitives}. These metrics
characterize the nature and magnitude of misspecification independent of
the agent, revealing which language ecosystems are structurally harder to
specify. Second, retrieval measures of precision, recall, and F1 consider
$D^{(3)}$ as the ground truth and characterize per-agent accuracy, enabling
the direct cross-agent comparison of Section~\ref{sec:agents}.

For a successful project $(a, p, \ell)$, let
$n_j = |D^{(j)}_{a,p,\ell}|$ for $j \in \{1,2,3\}$. The three mismatch rates are

\vspace{5pt}
\begin{equation}
\phi = \tfrac{|D^{(1)} \setminus D^{(3)}|}{n_1}, \;\;
\eta = \tfrac{|D^{(3)} \setminus D^{(1)}|}{n_3}, \;\;
\beta = \tfrac{|D^{(2)} \setminus D^{(3)}|}{n_2},
\label{eq:rates}
\end{equation}
with each rate set to zero when its denominator is zero. The phantom rate
$\phi$ measures over-specification: packages declared but never loaded at
runtime. The hidden rate $\eta$ measures under-specification: packages
loaded at runtime but absent from the initial manifest. The bloat rate
$\beta$ measures transitive excess: packages installed by the package
manager but never accessed during execution. Treating $D^{(1)}$ as a
retrieval system predicting $D^{(3)}$, per-project precision, recall, and
F1 are
\begin{equation}
\mathrm{P} = \frac{|D^{(1)} \cap D^{(3)}|}{n_1}, \;\;
\mathrm{R} = \frac{|D^{(1)} \cap D^{(3)}|}{n_3}, \;\;
\mathrm{F1} = \frac{2\,\mathrm{P}\,\mathrm{R}}{\mathrm{P} + \mathrm{R}},
\label{eq:prf}
\end{equation}
with the conventions $\mathrm{P} = 1$ when $n_1 = 0$ and $n_3 > 0$,
$\mathrm{R} = 1$ when $n_3 = 0$ and $n_1 > 0$, and projects with
$n_1 = n_3 = 0$ excluded entirely. Precision and recall jointly
characterize the direction of agent error: high precision but low recall
indicates systematic under-specification, while the reverse indicates
systematic over-specification. All measures are aggregated as means per
stratum $(a, \ell)$ and per ecosystem $\ell$ pooled over agents.

Figure~\ref{fig:m3_phb} shows the character of misspecification per
ecosystem. Python and Java have relatively low phantom rates (about 5\%),
but both still show meaningful hidden-dependency rates because transitive
packages become active at runtime without being named in the initial
manifest. JavaScript is much worse on all three dimensions: about 49\% of
declared packages are phantom, about 62\% of runtime packages are hidden,
and about 22\% of installed packages are never touched. C++ is dominated by
a different pathology: many dependencies are compile-time or header-only
and therefore disappear from the runtime trace, inflating phantom and bloat
rates under a runtime-ground-truth definition.

Figure~\ref{fig:m2_inflation} quantifies how large the eventual installed
environment becomes relative to what the agent initially declared, using
the inflation ratio $\rho$ defined in Section~\ref{sec:primitives}.
JavaScript shows by far the largest environment growth, with
$\bar{\rho}_{\mathrm{JavaScript}} = 11.40$ and a median of 8.5, meaning
that small initial declarations routinely lead to much larger installed
environments. Java also shows heavy inflation
($\bar{\rho}_{\mathrm{Java}} = 3.97$), while Python is noticeably milder
($\bar{\rho}_{\mathrm{Python}} = 2.06$). C++ remains near $\rho = 1.0$,
which matches its non-centralized dependency workflow
(Section~\ref{sec:primitives}), with only minor deviations in rare
fetched-project cases. The figure shows that the size of the initial
declared dependency set is a poor proxy for actual footprint in
resolver-heavy ecosystems.

Table~\ref{tab:manifest_accuracy} reports the retrieval measures. Python
achieves the strongest overall initial-manifest accuracy with macro F1 of
0.765, followed by Java at 0.708. JavaScript drops sharply to 0.219 because
both over-declaration and under-declaration are common. C++ falls to 0.018,
but that value should be interpreted together with the methodological
limitation that many C++ libraries leave little or no runtime footprint
after compilation (Section~\ref{sec:limitations}).

The impact of these mismatch rates extends beyond code quality. A dependency
manifest is the primary input to software bills of materials, vulnerability
scans, and license audits; when nearly half of a JavaScript manifest is
phantom and nearly two thirds of what actually runs is hidden, every
downstream audit inherits that inaccuracy. Phantom declarations inflate the
apparent attack surface and generate false positives in security review,
while hidden dependencies are worse: they are runtime-active code that no
scan ever examines. At the accuracy levels of
Table~\ref{tab:manifest_accuracy}, an agent-produced manifest cannot be
treated as a trustworthy audit artifact in JavaScript or C++ without
independent runtime verification.

% ─────────────────────────────────────────────────────────────────────
\subsection{Cross- and Intra-Agent Comparison}
\label{sec:agents}
% ─────────────────────────────────────────────────────────────────────

\begin{figure}[t]
\centering
\includegraphics[width=\columnwidth]{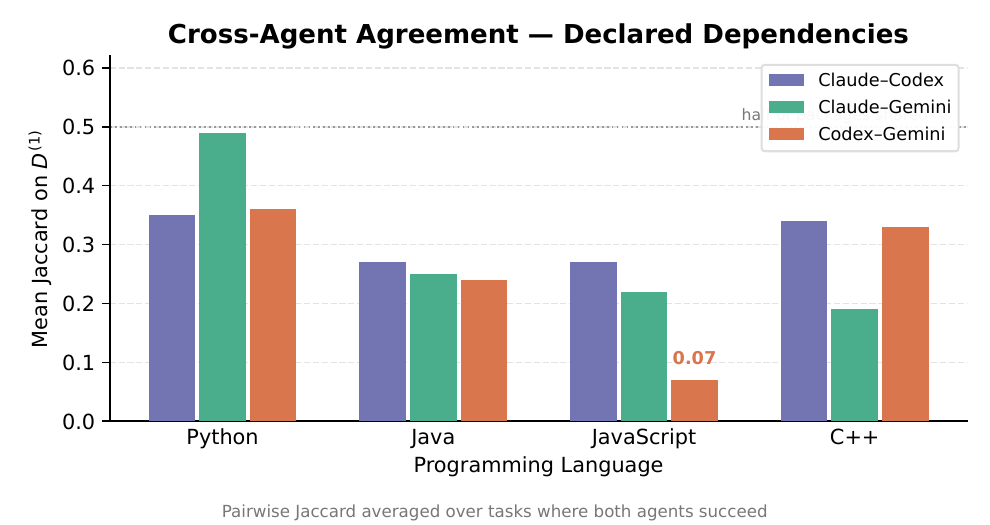}
\caption{Cross-agent agreement on declared dependencies. Each bar reports
the mean Jaccard similarity between the declared dependency sets extracted
from two agents' initial manifests for the same task in the same
ecosystem; the dotted line marks the level at which half of the declared
packages would be shared.}
\label{fig:m5_jaccard}
\end{figure}

\begin{table}[t]
\caption{Stochastic variability: Claude's three independent trials produce
near-zero agreement.}
\label{tab:stochastic}
\centering
\small
\begin{tabular}{lccccc}
\toprule
\rowcolor{HeadTint}
\textbf{Language} & \textbf{Mean $\bar{J}_{\mathrm{intra}}$} & \textbf{Median} & \textbf{UCR} & \textbf{$\bar{V}_{\cup}$} & \textbf{$\bar{V}_{\cap}$} \\
\midrule
  \textcolor{LPy}{\textbf{Python}} & 0.113 & 0.083 & 0/50 (0.0\%) & 5.5 & 0.2 \\
  \textcolor{LJava}{\textbf{Java}} & 0.138 & 0.111 & 0/48 (0.0\%) & 5.6 & 0.1 \\
  \textcolor{LJS}{\textbf{JavaScript}} & 0.068 & 0.000 & 0/50 (0.0\%) & 5.9 & 0.0 \\
  \textcolor{LCpp}{\textbf{C++}} & 0.093 & 0.067 & 0/35 (0.0\%) & 5.3 & 0.1 \\
\bottomrule
\end{tabular}
\end{table}

We now measure cross-agent and intra-agent agreement. For both assessments,
we define the pairwise declared-dependency agreement between agents $a_i$
and $a_j$ on task $p$ in language $\ell$ as
\begin{equation}
  J(a_i, a_j, p, \ell)
  \;=\;
  \frac{|D^{(1)}_{a_i} \cap D^{(1)}_{a_j}|}
       {|D^{(1)}_{a_i} \cup D^{(1)}_{a_j}|},
  \label{eq:jaccard}
\end{equation}
with $J = 1$ when both sets are empty and $J = 0$ when exactly one is. The
mean cross-agent Jaccard $\bar{J}_\ell$ and mean intra-agent Jaccard
$\bar{J}_{\mathrm{intra}}$ are averages of $J$ over all valid agent pairs
and, for the intra-agent case, over the three independent Claude Code
trials of Section~\ref{sec:setup}. The complete disjointness rate
$\mathrm{CDR}_\ell$ is the fraction of pairs for which $J = 0$, and the
unanimous consensus rate $\mathrm{UCR}_\ell$ is the fraction of tasks for
which all three Claude trials produce an identical $D^{(1)}$.

Figure~\ref{fig:m5_jaccard} shows that agents consistently solve the same
programming task with substantially different declared dependency sets.
Python is the most convergent ecosystem, yet even there mean pairwise
Jaccard similarity remains below 0.5 across all agent pairs, indicating
that fewer than half of declared packages are shared on average.
JavaScript is the most divergent, with Codex and Gemini agreeing on as
little as $\bar{J} = 0.073$, meaning the two agents share fewer than one in
thirteen declared packages for the same task. The complete disjointness
rate is highest in JavaScript (63.3\%), followed by C++ (57.7\%) and Java
(56.2\%), with Python lowest at 36.0\%, yielding a macro-average
$\mathrm{CDR}$ of 53.3\% across the four ecosystems. Complete disagreement
between agents is therefore not an edge case but the majority outcome.

Table~\ref{tab:stochastic} asks whether this divergence is vendor-specific
or whether it reflects a deeper instability, by repeating Claude Code three
independent times on the same tasks. The unanimous consensus rate is 0\% in
every language, meaning Claude never converged to the same dependency
declaration across all three trials for any task in any ecosystem.
Intra-agent Jaccard scores are low throughout, with JavaScript again the
worst case. Crucially, the mean vocabulary union $\bar{V}_{\cup}$
substantially exceeds the mean unanimous core $\bar{V}_{\cap}$ in every
language, confirming that the agent draws from a much larger pool of
plausible packages than it consistently commits to. Taken together, the
figure and table show that dependency selection is neither reproducible
across vendors nor stable within a single model, a result that implicates
the generative process itself rather than any vendor-specific design
choice, and suggests that dependency specification is not retrieved from a
stable internal knowledge structure but sampled stochastically on each
generation.

The practical consequence is that agent-generated software is not
reproducible at the supply-chain level: two developers prompting the same
agent with the same task, or one developer prompting it twice, receive
different dependency trees, so every fresh generation effectively produces
a different supply chain. Dependency review, pinning decisions, and
security approvals performed on one generation do not transfer to the
next. Because the instability persists within a single vendor, it also
cannot be mitigated by standardizing on one agent; it must be addressed in
the generation process itself.

% ─────────────────────────────────────────────────────────────────────
\subsection{Cross-Language Variation}
\label{sec:languages}
% ─────────────────────────────────────────────────────────────────────

\begin{table}[t]
\caption{Environment Gap: C++ system library assumption failures and
recovery. ``Fail'' counts first-attempt failures; ``SysLib'' counts those
whose terminal error is a missing system library; ``Rec.'' counts SysLib
failures eventually repaired.}
\label{tab:env_gap}
\centering
\small
\setlength{\tabcolsep}{4pt}
\begin{tabular}{@{}lccccc@{}}
\toprule
\rowcolor{HeadTint}
\textbf{Agent} & \textbf{Fail} & \textbf{SysLib} & \textbf{SLAR} & \textbf{Rec.} & \textbf{EGAR} \\
\midrule
  \textcolor{AgClaude}{\textbf{Claude}} & 32 & 28 & 87.5\% & 28 & 100.0\% \\
  \textcolor{AgCodex}{\textbf{Codex}}   & 28 &  5 & 17.9\% &  5 & 100.0\% \\
  \textcolor{AgGemini}{\textbf{Gemini}} & 10 &  0 &  0.0\% &  0 & -- \\
\bottomrule
\end{tabular}
\end{table}

The preceding results consistently separate the four ecosystems, and we now
draw the language-level picture together. Python is the easiest ecosystem
to specify: it combines the highest first-attempt success, the lowest
phantom rates, mild environment inflation
($\bar{\rho}_{\mathrm{Python}} = 2.06$), and the best initial-manifest
accuracy (macro F1 of 0.765). Java sits in the middle: first-attempt
success is weaker, but its declaration quality remains high (macro F1 of
0.708) despite heavy transitive inflation. JavaScript is the pathological
case on every specification metric: the largest inflation
($\bar{\rho} = 11.40$), the highest phantom and hidden rates, the lowest
manifest accuracy (0.219), and the weakest cross- and intra-agent
agreement. Poor specification quality is therefore not merely an agent
property; it interacts strongly with how aggressively an ecosystem's
package manager expands and obscures the declared set.

C++ varies along a different axis: agents fail not because the resolver
inflates their declarations, but because they assume system development
libraries that the clean container does not provide. For every C++ project
that fails at round $i = 0$, we classify the terminal error $e^{(0)}$ into
one of three types: \textbf{SystemLib}, a \texttt{find\_package} failure
for a system library absent from $\mathcal{D}_\ell$ (e.g.,
\texttt{OpenSSL}, \texttt{libcurl}, \texttt{SQLite3});
\textbf{FetchFail}, a \texttt{FetchContent\_Declare} network or archive
error; and \textbf{Other}, covering compilation, missing-header, linker, or
runtime failures unrelated to system library absence. From this taxonomy we
report two rates per agent: the \emph{system library assumption rate}
(SLAR), the fraction of an agent's first-attempt C++ failures whose
terminal error is of type SystemLib, and the \emph{Environment Gap
recovery rate} (EGAR), the fraction of those SystemLib failures that the
agent eventually repairs within the budget.

Table~\ref{tab:env_gap} reports the clearest C++ manifestation of the
environment gap. Claude's initial C++ failures are overwhelmingly due to
incorrect system-library assumptions (SLAR of 87.5\%), and every one of
those failures is eventually repaired after feedback. Codex shows a lower
but still meaningful assumption rate (17.9\%), again with perfect recovery
once the missing package is revealed. Gemini's C++ failures arise from
other causes rather than missing system libraries. The table shows that
the clean Docker image is not just making the benchmark harder; it is
exposing a concrete mismatch between the environments agents implicitly
assume and the environments they are actually asked to run in.

Together, these ecosystem differences imply that mitigation cannot be
uniform. JavaScript projects need the strongest specification guardrails,
because there the manifest is least trustworthy precisely where the
resolver amplifies it most. C++ points to a different remedy: since nearly
all system-library assumption failures are repaired once the environment
reports the missing package, feedback from the actual deployment target,
rather than a richer assumed environment, is the effective corrective. Code
that runs on a developer's fully provisioned machine will still fail in the
minimal containers that production systems favor, and the environment gap
measures exactly that deployment risk.

% ─────────────────────────────────────────────────────────────────────
\subsection{Task Complexity Sensitivity}
\label{sec:complexity}
% ─────────────────────────────────────────────────────────────────────
\begin{figure}[t]
\centering
\includegraphics[width=\columnwidth]{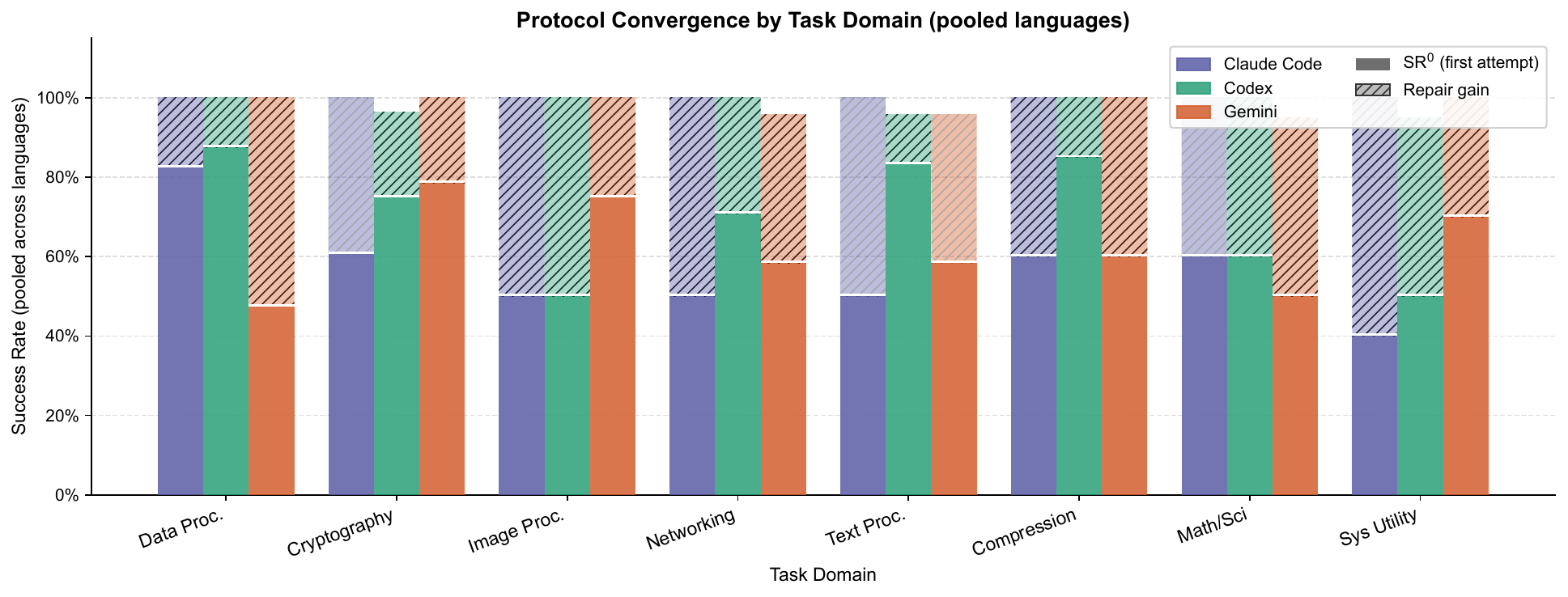}
\caption{Protocol convergence by task domain, with solid and hatched bars
as in Figure~\ref{fig:conv_language}.}
\label{fig:conv_domain}
\end{figure}
Finally, we ask whether specification quality depends on what kind of software is being built. Figure~\ref{fig:conv_domain} repeats the convergence analysis of Section~\ref{sec:conv} per task domain. Final success is at or near 100\% in every domain (never below 95\% for any agent--domain pair), and no domain is intrinsically hard for all agents: first-attempt success spans 40\%--88\% across domain--agent pairs, but the weakest domain differs per agent --- System Utilities for Claude (40\%), Image Processing for Codex (50\%), and Data Processing for Gemini (48\%), which is simultaneously the strongest domain for the other two agents. Convergence behavior is therefore driven primarily by language ecosystem rather than by task domain: C++ is hard for every agent regardless of what the task computes.

A sharper test of task sensitivity isolates the tasks that need no external libraries at all. Following Section~\ref{sec:setup}, a subset of benchmark tasks in each language is identified \emph{a priori} by the authors as entirely solvable with the standard library (e.g., Python's \texttt{csv}, \texttt{hashlib}, and \texttt{json} modules; Java's \texttt{java.util} and \texttt{java.security} packages; Node's \texttt{fs}, \texttt{crypto}, and \texttt{path} built-ins). The \emph{unnecessary dependency rate} (UDR) is the fraction of these standard-library-sufficient tasks for which the agent nevertheless declared at least one external package.

On these standard-library-sufficient tasks, Claude introduced external packages in 91.8\% of these tasks on macro average, reaching 100\% in C++ and remaining at or above 84\% in every other ecosystem. Gemini is similarly aggressive in JavaScript. Codex is much more conservative in Python and JavaScript, though it still adds unnecessary libraries in more than half of eligible C++ tasks. This result isolates over-dependence from task difficulty: the extra libraries are not required by the benchmark itself.

Each unnecessary package is avoidable attack surface, maintenance burden, and audit scope added for zero functional gain, and on standard-library-sufficient tasks that cost is pure waste. Because the behavior persists on tasks where no library is needed, it reflects a generation prior rather than a response to task demands, which makes it a natural target for training objectives that reward minimal, sufficient environment specification alongside functional correctness.

% ─────────────────────────────────────────────────────────────────────
\subsection{Limitations}
\label{sec:limitations}
% ─────────────────────────────────────────────────────────────────────
The strongest measurement limitation is in C++. Runtime tracing undercounts
header-only and purely compile-time dependencies, which pushes C++ manifest
precision and F1 downward relative to ecosystems where runtime loading
leaves a clearer file-level footprint. Similarly, although C++ usually has
$|D^{(2)}| \approx |D^{(1)}|$, a small number of fetched projects
recursively pull an additional library during their own build process, so
exact equality should be interpreted as an ecosystem-level trend rather
than as a strict invariant. More broadly, the study uses a controlled
benchmark of 50 tasks rather than fully open-ended real-world projects, so
the reported rates should be interpreted as evidence about structured
code-generation settings rather than as exhaustive estimates for all
software development contexts. All results are additionally conditioned on
the prompt template of Section~\ref{sec:prompt}; because the template is
held identical across agents and languages, comparative findings are
controlled at the prompt level, but absolute rates may vary under
different phrasings, and a systematic prompt-robustness analysis remains
future work. The manual evaluation protocol and fixed annotations for
standard-library tasks and C++ failure labels improve control, but they
also introduce a degree of judgment that future larger-scale replications
should further test. Finally, because commercial coding agents evolve
rapidly, the results should be understood as a snapshot of the evaluated
agent configurations at evaluation time rather than as immutable model
properties.

%% file: sections/07_related.tex
% ═══════════════════════════════════════════════════════════════════════
\section{Related Work}
\label{sec:related}

\vspace{5pt}
\noindent \textbf{LLM Code Generation and Its Evaluation Gaps.} The era of LLM code generation began with Codex~\citep{chen2021codex}, which solved 28.8\% of HumanEval problems. Subsequent models such as AlphaCode~\citep{li2022alphacode}, StarCoder~\citep{li2023starcoder}, Code Llama~\citep{roziere2024codellama}, and commercial agents like Claude Code~\citep{anthropic2024claude}, GitHub Copilot, and Gemini Code Assist~\citep{team2024gemini} have expanded from single-function completion to full project synthesis. 
Surveys by \citet{fan2023large} and \citet{jiang2024survey} cover this landscape comprehensively.

The critical gap in this literature is that nearly all benchmarks evaluate functional correctness of isolated snippets~\citep{chen2021codex, austin2021mbpp}. A smaller body of work examines code quality~\citep{liu2024refining}, bug patterns~\citep{jesse2023large, tambon2024bugs}, and security vulnerabilities~\citep{pearce2022asleep}. \citet{ouyang2023llm} studied non-determinism in ChatGPT's code output but restricted their analysis to functional behavior; they did not examine whether dependency choices are stable across runs. Our stochastic analysis fills this gap directly: we show that a single agent's dependency declarations are unstable across independent runs of the same task, with unanimous consensus rate zero across all tasks and languages — a form of non-determinism invisible to functional correctness metrics. Recent work has identified \emph{package hallucinations}, where models recommend packages that do not exist in registries, linked to practical supply-chain attacks~\citep{lanyado2023hallucination}. Our work is complementary but distinct: we study the misspecification of \emph{existing} packages,  over-declaration, under-declaration, and transitive blindness, which is a different and broader failure mode that persists even when every named package exists. None of this work examines dependency management as a first-class evaluation concern across the full lifecycle from declaration through resolution to runtime execution. Our own earlier study~\citep{vangala2026aigenerated} provided initial evidence of dependency gaps in agent-generated code; the present article formalizes the evaluation protocol underlying that study and extends the analysis.

\noindent \textbf{Dependency Ecosystems and Their Distinct Challenges.}
Software dependency management is not a uniform problem: each ecosystem has its own resolution semantics, transitive behavior, and failure modes, and these differences matter profoundly for LLM-generated code. A key contribution of this paper is measuring these ecosystem-specific failure modes under a unified framework, rather than studying any single ecosystem in isolation. We breifly review relevant works within each ecosystem: 

\paragraph{JavaScript/npm.} \citet{zimmermann2019small} found that installing a median npm package transitively imports 79 packages maintained by 39 individuals. The npm ecosystem's deeply nested dependency graphs mean that a single declared package can pull in dozens or even hundreds of transitive dependencies, creating a vast and largely invisible attack surface. \citet{decan2019empirical} confirmed that npm exhibits the most aggressive growth and fragility among seven ecosystems studied. Our results quantify this aggression in the context of agent-generated code specifically: we observe a mean inflation ratio $\bar{\rho} = 11.40$ for JavaScript, with individual projects expanding a three-package declaration into over 140 installed packages, a transitive footprint no agent explicitly reasoned about.

\paragraph{Java/Maven.} \citet{soto2021comprehensive} found that 75.1\% of Maven Central artifacts contain at least one bloated dependency --- a library declared in the build file but never used at runtime. Maven's transitive resolution is opaque: a developer declares a handful of dependencies in \texttt{pom.xml}, and Maven silently resolves an entire tree of transitive artifacts. \citet{wang2020empirical} further documented the risks of third-party library management in Java projects. Whereas these studies measure bloat in human-authored artifacts, we measure it in agent-generated code where the selection decisions are made by a model rather than a developer, finding that bloat is pervasive and that agents show no awareness of the transitive footprint their declarations induce.

\paragraph{C++/CMake.} C++ dependency management is fundamentally different from the other three ecosystems. There is no single dominant package manager; dependencies are handled through CMake's \texttt{FetchContent}, \texttt{find\_package}, or manual vendoring. The key issue for LLM-generated code is the \emph{system library assumption gap}: agents trained on code from rich development environments assume that system libraries such as \texttt{libssl-dev} or \texttt{libopencv-dev} are already installed. We provide the first systematic measurement of this assumption gap across multiple agents, showing that it accounts for the majority of C++ build failures and that its prevalence differs substantially across agents — implicating training distribution rather than task difficulty as the primary driver.

\paragraph{Python/pip.} Python's ecosystem is comparatively well-behaved for dependency management. The \texttt{pip freeze} convention produces exact version pins, and transitive resolution is relatively flat. However, system library issues still arise for packages with native extensions --- \texttt{pyzbar} requires \texttt{libzbar0} and \texttt{opencv-python} requires \texttt{libGL.so} --- and our results show that even in this most tractable ecosystem, agents achieve macro F1 of only 0.765 against runtime ground truth, leaving meaningful specification gaps.

\vspace{5pt}
\noindent \textbf{Cross-ecosystem analysis.} \citet{kikas2017structure} studied the structural evolution of dependency networks in JavaScript, Ruby, and Rust. \citet{cox2019surviving} articulated the fundamental tension between code reuse and dependency risk. Supply-chain attacks exploiting dependencies are a major concern~\citep{ladisa2023taxonomy, liu2022demystifying}, prompting government mandates for Software Bills of Materials~\citep{ntia2021sbom, whitehouse2021executive}. Our paper provides the first cross-ecosystem comparison of dependency specification quality in agent-generated code, using a unified three-layer measurement framework that enables direct comparison across Python, Java, JavaScript, and C++ rather than treating each ecosystem as an isolated problem.

\vspace{5pt}
\noindent \textbf{Software Bloat and Unused Dependencies.}
\citet{bruce2020jshrink} found 14.8\% of compiled Java methods unreachable. \citet{quach2018debloating} proposed piece-wise compilation for C/C++ attack surface reduction. \citet{soto2023coverage} showed that most transitive Maven dependencies have very low runtime utilization. These studies measure bloat in human-authored artifacts using static or coverage-based analysis. Our study is the first to measure dependency bloat in agent-generated code, where the selection decisions are made by a model rather than a developer, and to use dynamic runtime tracing as the ground truth rather than static reachability, capturing the actual runtime footprint rather than an approximation of it.

\vspace{5pt}
\noindent \textbf{Static vs Dynamic Dependency Analysis.}
Tools for automatically inferring dependencies from source code, such as \texttt{pipreqs} for Python, \texttt{jdeps} for Java, and import-graph analyzers more broadly, represent the non-agent baseline for dependency specification. These tools use static analysis, which has known limitations: they cannot capture dynamically imported packages, conditionally loaded modules, or runtime-only dependencies exercised on specific execution paths. Dynamic tracing via system call interposition addresses these limitations at the cost of execution-path specificity — capturing only dependencies exercised during a particular run. We use dynamic tracing via \texttt{ptrace}-based provenance capture~\citep{chirigati2016reprozip, tonthat2017sciunit} as our runtime ground truth precisely because it captures what the program actually loads. %, not what a static analysis predicts it might load, while acknowledging execution-path specificity as a scope limitation. %The contrast between $D^{(1)}$ (agent declaration), $D^{(2)}$ (package manager resolution), and $D^{(3)}$ (runtime trace) is designed to isolate the contribution of each analysis method to the total dependency picture.

\vspace{5pt}
\noindent \textbf{Software Bills of Materials and Manifest Accuracy.}
Regulatory mandates for Software Bills of Materials~\citep{ntia2021sbom, whitehouse2021executive} have prompted a growing literature on SBOM generation accuracy and completeness~\citep{zahan2023software, bi2023empirical}. Existing SBOM generation tools operate on human-authored codebases; the accuracy of agent-generated manifests as SBOMs has not been studied. Our three-layer framework is directly applicable to this question: the declared dependency set $D^{(1)}$ is precisely what an agent-generated SBOM would contain, and our precision, recall, and F1 measurements against runtime ground truth $D^{(3)}$ provide the first empirical characterization of agent-generated SBOM accuracy. The finding that macro F1 ranges from 0.765 for Python to 0.018 for C++ has direct implications for the trustworthiness of agent-generated SBOMs in security-critical contexts.

\vspace{5pt}
\noindent \textbf{Computational Reproducibility and Provenance.}
Reproducibility remains a persistent challenge in computational science~\citep{baker2016reproducibility, collberg2016repeatability}. Docker containers~\citep{boettiger2015introduction} capture static environment snapshots but cannot reveal what a program actually loads at runtime — a limitation that makes them insufficient as a ground truth for dependency analysis. Provenance-based systems go further. ReproZip~\citep{chirigati2016reprozip} traces system calls to build reproducible archives. Sciunit~\citep{pham2013using} uses its PTU engine  building on CDE's system call interposition approach~\citep{guo2011cde} to intercept \texttt{open()}, \texttt{openat()}, \texttt{read()}, and \texttt{execve()} via \texttt{ptrace}, recording every file accessed during execution. We use Sciunit as our runtime dependency oracle, extending its application from reproducibility packaging to dependency ground-truth extraction in the context of agent-generated code evaluation.

\vspace{5pt}
\noindent \textbf{LLM Evaluation Methodology.}
Beyond code generation specifically, our work contributes to the broader literature on LLM evaluation methodology. Holistic evaluation frameworks such as HELM~\citep{liang2022holistic} argue that model capabilities should be measured across multiple dimensions simultaneously rather than through single-metric proxies. Our three-layer dependency framework instantiates this principle for the code generation setting: rather than collapsing environment specification quality into a single pass/fail execution outcome, we decompose it into declaration accuracy, resolution behavior, and runtime footprint, each of which implicates a different failure mode and a different remediation. We argue that environment specification should be incorporated as a standard evaluation dimension in code generation benchmarks alongside functional correctness, and that the metrics introduced here - phantom rate, hidden rate, bloat rate, and manifest F1 - provide a reusable evaluation vocabulary for this dimension.

%% file: sections/08_conclusion.tex
% ═══════════════════════════════════════════════════════════════════════
\section{Conclusion}
\label{sec:conclusion}
% ═══════════════════════════════════════════════════════════════════════
This study shows that dependency generation by current LLM coding agents is
highly inconsistent and often misaligned with the environments their code
actually requires. Although these agents can frequently repair projects to
functional success, they do not reliably produce correct or minimal dependency
specifications.
Our results directly answer the central research question: current LLMs
systematically misspecify software environment dependencies, and this
misspecification varies substantially across ecosystems and across agents. The
problem appears both as under-specification, where required dependencies are
missing from the manifest, and as over-specification, where unnecessary
libraries inflate the software environment.
%Several quantitative findings make this clear. JavaScript exhibits the largest
% environment growth, with a mean growth ratio of
% $\bar{\rho}=11.40$, while Python and Java remain substantially lower at 2.06
% and 3.97, respectively. Initial-manifest accuracy is strongest in Python
% (macro F1\,=\,0.765) and Java (0.708), but degrades sharply in JavaScript
% (0.219) and becomes near-zero in C++ (0.018) under runtime-ground-truth
% evaluation. Cross-agent agreement is also weak: complete disjointness occurs in
% 53.3\% of identical-task comparisons, and Claude's three-run stochastic
% analysis yields 0\% unanimous consensus in every ecosystem. Even on tasks
% solvable using only the standard library, Claude introduces external
% dependencies in 91.8\% of cases on macro average.
% These results suggest that dependency selection is often heuristic rather than
% consistently aligned with the actual runtime requirements of the program.
% Instead, dependency choice appears to be ecosystem-sensitive and strongly
% shaped by training-time assumptions about what is already available in the
% environment. In this sense,
% dependency specification is not a minor implementation detail, but a distinct
% failure mode of current code-generation systems.
These findings have immediate implications for both research and practice.
Successful code generation should not be treated as evidence of environment
correctness without additional validation. More reliable and portable code
generation will require evaluation frameworks, validation pipelines, and
training objectives that explicitly target accurate and minimal dependency
specification alongside task completion.
One concrete direction is to treat dependency specification as a first-class
learning target: future systems could incorporate environment-grounded
fine-tuning, dependency-aware reward signals, or manifest-verification steps in
iterative generation and repair loops so that dependency correctness is
optimized alongside functional task completion. %This framing turns dependency
% specification from a downstream tooling issue into a trainable and evaluable
% target for learning-based code generation.
% Ultimately, reliable software generation requires not only correct code, but
% correct dependency reasoning.

%% file: sections/09_appendix.tex
% =======================================================================
% APPENDIX
% =======================================================================
\appendix

% ─────────────────────────────────────────────────────────────────────
\section{Supplementary Structural Metrics}
\label{sec:appendix}
% ─────────────────────────────────────────────────────────────────────

This appendix reports supplementary results that support interpretation of
the primary findings.

\paragraph{Manifest size vs.\ first-attempt success.}
We test whether projects with larger declared dependency sets fail more
frequently on their first execution attempt. For each ecosystem $\ell$, we
compute the Spearman rank correlation between $|D^{(1)}_{a,p,\ell}|$ and
the round-zero success indicator
$\mathbf{1}[\mathrm{Success}(C^{(0)}, M^{(0)}, \mathcal{D}_\ell)]$ over
all $(a, p)$ pairs. A negative correlation would indicate that manifest
bloat is itself a source of build fragility.

\begin{table}[htbp]
\caption{Spearman correlation between manifest size and first-attempt
success.}
\label{tab:spearman_manifest_success}
\centering
\small
\begin{tabular}{@{}lcc@{}}
\toprule
\rowcolor{HeadTint}
\textbf{Ecosystem} & \textbf{$\rho_S$} & \textbf{$p$-value} \\
\midrule
\textcolor{LPy}{\textbf{Python}}     & $-0.276$ & 0.0006 \\
\textcolor{LJava}{\textbf{Java}}     & $-0.260$ & 0.0013 \\
\textcolor{LJS}{\textbf{JavaScript}} & $-0.093$ & 0.2564 \\
\textcolor{LCpp}{\textbf{C++}}       & $-0.410$ & 0.0000 \\
\bottomrule
\end{tabular}
\end{table}

Table~\ref{tab:spearman_manifest_success} shows that larger declared
dependency sets are associated with lower first-attempt success in Python,
Java, and C++. The strongest negative association appears in C++, while
JavaScript shows the same direction but not a statistically meaningful
effect in the current benchmark.

\paragraph{Summary.}
Table~\ref{tab:metrics} cross-references the eight primary metrics against
the results subsections where they are reported.

\begin{table}[htbp]
\caption{Metric summary: definitions, key quantities, and result section.}
\label{tab:metrics}
\centering
\normalsize
\setlength{\tabcolsep}{4pt}
\begin{tabularx}{\columnwidth}{@{}l>{\raggedright\arraybackslash}Xll@{}}
\toprule
\rowcolor{HeadTint}
\textbf{ID} & \textbf{Metric} & \textbf{Key quantity} & \textbf{Results} \\
\midrule
M1 & Success rate under self-correction
   & $\mathrm{SR}^{(0)}$, $\mathrm{SR}^{(\leq k)}$
   & \S\ref{sec:conv} \\
M2 & Environment growth
   & $\rho$
   & \S\ref{sec:quality} \\
M3 & Phantom / hidden / bloat rates
   & $\phi$, $\eta$, $\beta$
   & \S\ref{sec:quality} \\
M4 & Manifest accuracy
   & P, R, F1
   & \S\ref{sec:quality} \\
M5 & Cross-agent consistency
   & $\bar{J}_\ell$, CDR
   & \S\ref{sec:agents} \\
M6 & Stochastic variability
   & $\bar{J}_{\mathrm{intra}}$, UCR
   & \S\ref{sec:agents} \\
M7 & Unnecessary external libraries
   & UDR
   & \S\ref{sec:complexity} \\
M8 & Environment gap failures
   & SLAR, EGAR
   & \S\ref{sec:languages} \\
\bottomrule
\end{tabularx}
\end{table}

% ─────────────────────────────────────────────────────────────────────
\section{Complete Problem List}
\label{app:tasks}
% ─────────────────────────────────────────────────────────────────────

\begin{table*}[htbp]
\caption{All 50 programming problems.}
\label{tab:all_tasks}
\centering
\scriptsize
\setlength{\tabcolsep}{4pt}
\begin{tabular}{@{}cllc@{\hspace{14pt}}cllc@{}}
\toprule
\rowcolor{HeadTint}
\textbf{ID} & \textbf{Problem} & \textbf{Domain} & \textbf{Cx.} &
\textbf{ID} & \textbf{Problem} & \textbf{Domain} & \textbf{Cx.} \\
\midrule
p\_01 & CSV Statistical Analyzer     & Data Processing & M & p\_26 & DNS Resolver                 & Networking       & M \\
p\_02 & JSON Schema Validator        & Data Processing & M & p\_27 & HTTP REST Client w/ Retry    & Networking       & H \\
p\_03 & XML to JSON Converter        & Data Processing & M & p\_28 & Static HTTP File Server      & Networking       & H \\
p\_04 & YAML Config Merger           & Data Processing & M & p\_29 & Network Latency Monitor      & Networking       & M \\
p\_05 & Log File Pattern Analyzer    & Data Processing & M & p\_30 & Markdown to HTML Converter   & Text Processing  & M \\
p\_06 & SQLite Database Manager      & Data Processing & H & p\_31 & Spell Checker                & Text Processing  & M \\
p\_07 & Data Profiling Tool          & Data Processing & H & p\_32 & TF-IDF Search Engine         & Text Processing  & H \\
p\_08 & Multi-Format Data Converter  & Data Processing & H & p\_33 & Text Diff Tool               & Text Processing  & M \\
p\_09 & Time Series Trend Detector   & Data Processing & H & p\_34 & HTML to Plain Text           & Text Processing  & M \\
p\_10 & Duplicate Record Finder      & Data Processing & M & p\_35 & Template Engine              & Text Processing  & H \\
p\_11 & AES-256 File Encryption      & Cryptography    & H & p\_36 & ZIP Archive Manager          & Compression      & M \\
p\_12 & RSA Digital Signatures       & Cryptography    & H & p\_37 & GZIP Compression Tool        & Compression      & M \\
p\_13 & HMAC Integrity Checker       & Cryptography    & M & p\_38 & Tar Archive Handler          & Compression      & M \\
p\_14 & TOTP Generator               & Cryptography    & M & p\_39 & File Deduplicator            & Compression      & M \\
p\_15 & Password Strength Analyzer   & Cryptography    & M & p\_40 & Compression Benchmark        & Compression      & H \\
p\_16 & X.509 Certificate Parser     & Cryptography    & H & p\_41 & Matrix Operations Calculator & Math/Scientific  & H \\
p\_17 & Bcrypt Password Hasher       & Cryptography    & M & p\_42 & Linear Regression Fitter     & Math/Scientific  & H \\
p\_18 & Image Format Converter       & Image Processing & M & p\_43 & Dijkstra Shortest Path      & Math/Scientific  & M \\
p\_19 & Image Resizer / Thumbnails   & Image Processing & M & p\_44 & FFT Spectrum Analyzer       & Math/Scientific  & H \\
p\_20 & Image Watermarking Tool      & Image Processing & M & p\_45 & Numerical Integration       & Math/Scientific  & H \\
p\_21 & Image Histogram Analyzer     & Image Processing & M & p\_46 & QR Code Generator / Reader  & System Utilities & M \\
p\_22 & EXIF Metadata Extractor      & Image Processing & M & p\_47 & PDF Report Generator        & System Utilities & H \\
p\_23 & Color Palette Extractor      & Image Processing & H & p\_48 & Merkle Tree Verifier        & System Utilities & M \\
p\_24 & HTTP File Downloader         & Networking      & M & p\_49 & Cron Expression Parser      & System Utilities & M \\
p\_25 & TCP Port Scanner             & Networking      & M & p\_50 & Structured Log Processor    & System Utilities & M \\
\bottomrule
\end{tabular}
\vspace{2pt}

{\scriptsize Cx.\ = complexity: M = Medium, H = High.}
\end{table*}

\paragraph{Most frequent phantom and hidden packages.}
We compile ranked frequency tables of the packages that appear most often
in $D^{(1)} \setminus D^{(3)}$ (phantom) and $D^{(3)} \setminus D^{(1)}$
(hidden) across all projects in each ecosystem. These tables serve as
qualitative diagnostics: recurring phantom packages such as
\texttt{pycparser}, \texttt{yargs}, and \texttt{nlohmann\_json} reveal
systematic over-declaration patterns, while recurring hidden packages such
as \texttt{six}, \texttt{python-dateutil}, \texttt{jackson-core}, and
\texttt{jackson-annotations} reveal systematic omissions in dependency
reasoning.

\begin{table*}[htbp]
\caption{Most frequent phantom packages $D^{(1)} \setminus D^{(3)}$ by
ecosystem.}
\label{tab:phantom_packages}
\centering
\normalsize
\setlength{\tabcolsep}{4pt}
\begin{tabularx}{\columnwidth}{@{}l>{\raggedright\arraybackslash}X@{}}
\toprule
\rowcolor{HeadTint}
\textbf{Ecosystem} & \textbf{Most frequent phantom packages} \\
\midrule
\textcolor{LPy}{\textbf{Python}}     & \texttt{pycparser} (6), \texttt{pandas} (5), \texttt{tabulate} (2), \texttt{tzdata} (1), \texttt{python-dateutil} (1) \\[2pt]
\textcolor{LJava}{\textbf{Java}}     & \texttt{gson} (2), \texttt{zxing-javase} (2), \texttt{zxing-core} (2), \texttt{slf4j-api} (1), \texttt{javalin} (1) \\[2pt]
\textcolor{LJS}{\textbf{JavaScript}} & \texttt{yargs} (46), \texttt{sharp} (8), \texttt{fast-csv} (4), \texttt{lodash} (4), \texttt{js-yaml} (3) \\[2pt]
\textcolor{LCpp}{\textbf{C++}}       & \texttt{nlohmann\_json} (65), \texttt{openssl} (16), \texttt{opencv} (11), \texttt{yaml-cpp} (5), \texttt{mbedtls} (5) \\
\bottomrule
\end{tabularx}
\end{table*}

\begin{table*}[htbp]
\caption{Most frequent hidden packages $D^{(3)} \setminus D^{(1)}$ by
ecosystem.}
\label{tab:hidden_packages}
\centering
\normalsize
\setlength{\tabcolsep}{4pt}
\begin{tabularx}{\columnwidth}{@{}l>{\raggedright\arraybackslash}X@{}}
\toprule
\rowcolor{HeadTint}
\textbf{Ecosystem} & \textbf{Most frequent hidden packages} \\
\midrule
\textcolor{LPy}{\textbf{Python}}     & \texttt{six} (31), \texttt{python-dateutil} (23), \texttt{pytz} (20), \texttt{packaging} (14), \texttt{kiwisolver} (11) \\[2pt]
\textcolor{LJava}{\textbf{Java}}     & \texttt{jackson-annotations} (21), \texttt{jackson-core} (20), \texttt{slf4j-api} (14), \texttt{commons-logging} (12), \texttt{error\_prone\_annotations} (12) \\[2pt]
\textcolor{LJS}{\textbf{JavaScript}} & \texttt{ansi-styles} (15), \texttt{supports-color} (13), \texttt{has-flag} (13), \texttt{emoji-regex} (9), \texttt{is-fullwidth-}\allowbreak\texttt{code-point} (9) \\[2pt]
\textcolor{LCpp}{\textbf{C++}}       & \texttt{crypto} (16), \texttt{z} (14), \texttt{zstd} (6), \texttt{nettle} (6), \texttt{krb5} (5) \\
\bottomrule
\end{tabularx}
\end{table*}